\documentclass[
  aps,
  pra,
  onecolumn,
  10pt,
  amsmath,
  amssymb,
  floatfix,
  nofootinbib,
  superscriptaddress,
  longbibliography
]{revtex4-2}

\usepackage{amsthm}
\usepackage{physics}
\usepackage{graphicx}
\usepackage{subcaption}
\usepackage{booktabs}
\usepackage{enumitem}
\usepackage{xcolor}
\usepackage{mathtools}
\usepackage{hyperref}

\newcommand{\CE}{\mathrm{CE}}
\newcommand{\GCE}{\mathrm{GCE}}
\newcommand{\ee}{e}
\newcommand{\nbar}{\bar{n}}

\begin{document}

\title{Canonical-ensemble scaling of Tan's contact in the trapped Tonks--Girardeau gas}

\author{Felipe Taha Sant'Ana}
\affiliation{Arc$\hbar$imedian, 13560-120, São Carlos, Brazil}
\affiliation{Center of Mathematics, Computing, and Cognition, Federal University of ABC, 09210-580, Santo Andr\'e, Brazil}
\email{t.felipe@ufabc.edu.br}

\date{\today}

\begin{abstract}
We derive the canonical-ensemble scaling of Tan's contact for $N$
harmonically trapped Tonks--Girardeau bosons at finite temperature in
the large-$N$ limit. The leading scaling coefficient  reproduces the
local-density-approximation result and is obtained from a
contour-integral representation of the canonical partition function
followed by a saddle-point reduction to a phase-space integral with a
self-consistent scaled chemical potential. The subleading coefficient
is the central new object of this work: it admits an
explicit representation in terms of universal phase-space integrals
of the Fermi factor, having closed-form Sommerfeld  and
virial limits, and is identified with the
canonical-versus-grand-canonical ensemble difference at fixed mean
particle number. In the high-temperature Boltzmann regime the ratio
of subleading to leading coefficients collapses to a universal
value, traceable to the Poissonian
particle-number statistics of the dilute grand-canonical gas. We
construct Pad\'e approximants for both scaling functions that
interpolate uniformly between the low-temperature Sommerfeld and
high-temperature virial regimes; for the subleading coefficient we report a
form that is uniformly accurate on
our working range of temperatures and asymptotically correct beyond. The scaling law
is verified against canonical contour-integration data across the
full temperature range, and the experimental visibility of the
ensemble correction is quantified: for the mesoscopic particle
numbers and temperatures of current one-dimensional experiments it
reaches the several-percent level.
\end{abstract}

\maketitle

\section{Introduction}
%==========================================%==========================================
The momentum distribution of strongly interacting quantum gases exhibits
a universal $k^{-4}$ tail at large momentum, with a single
proportionality constant---Tan's contact $\mathcal{C}$---that controls
short-range pair correlations, the interaction energy, the adiabatic
sweep theorem and a host of related thermodynamic identities
\cite{Tan2008a,Tan2008b,Tan2008c}. Subsequent derivations of these
universal relations using the operator product expansion
\cite{BraatenPlatter2008}, generalisations to identical bosons
\cite{BraatenKangPlatter2011}, and extensions to general dimension and
spin \cite{ZhangLeggett2009,WernerCastin2012a,WernerCastin2012b,Combescot2009}
have established the contact as a unifying observable across the
strongly interacting cold-atom landscape, with experimental verifications
in three-dimensional Fermi gases \cite{Stewart2010,Kuhnle2010,Sagi2012}
and in atomic Bose-Einstein condensates \cite{Wild2012}.

One-dimensional (1D) Bose systems occupy a distinguished place in this
landscape because of integrability. The system composed of bosons
repulsively interacting through a $\delta$-function term is described 
by the Lieb--Liniger model \cite{LiebLiniger1963,Lieb1963}, which remains exactly solvable 
as long as no external potential is considered, where its finite-temperature
thermodynamics is governed by the Yang--Yang equations \cite{YangYang1969}.
In the limit of infinite repulsion the Lieb--Liniger gas reduces to the
Tonks--Girardeau (TG) gas, whose hard-core bosons are mapped onto free
spinless fermions by the Girardeau construction \cite{Girardeau1960};
yet the TG momentum distribution differs sharply from the Fermi one and
exhibits a non-trivial $k^{-4}$ contact tail. Quasi-one-dimensional
geometries are routinely realised by tight transverse confinement and
the associated confinement-induced resonance \cite{Olshanii1998}, which
provides the experimental gateway to the strongly interacting regime;
broader context can be found in the review \cite{Cazalilla2011}.

The TG and Lieb--Liniger regimes have been the subject of an extensive
experimental program. Direct observation of TG correlations was
reported in optical lattices \cite{Paredes2004} and in atom-chip and
crossed-dipole arrangements \cite{Kinoshita2004,Kinoshita2005,Haller2009},
followed by precision thermometry of 1D Bose gases
\cite{Jacqmin2011,Vogler2013}. More recent experiments have probed
strongly interacting 1D dynamics, including dynamical fermionisation
\cite{Wilson2020} and generalised hydrodynamics
\cite{Malvania2021}; very recently, the contact itself has been
measured directly in a 1D Lieb--Liniger gas \cite{Huang2025}, providing
a direct quantitative target for theoretical predictions of
the contact in terms of the number of particles and the temperature
in the trapped geometry.

On the theoretical side, the $k^{-4}$ tail of the trapped TG gas was
identified two decades ago \cite{Minguzzi2002}, with
the corresponding short-distance Lieb--Liniger analysis provided by \cite{OlshaniiDunjko2003}; the 1D version of Tan's
relations has also been systematised
\cite{BarthZwerger2011}, and the homogeneous finite-temperature pair
correlations were obtained from Yang--Yang thermodynamics by \cite{Kheruntsyan2003}. 
For the harmonically trapped TG gas at finite temperature, 
\cite{VignoloMinguzzi2013} derived a universal scaling through the  
local density approximation (LDA) within the grand-canonical-ensemble (GCE). Subsequent
work has extended this scaling to the trapped Lieb--Liniger gas
\cite{Yao2018,Lang2017} and to multi-component fermionic mixtures
\cite{Decamp2016,Decamp2018}, while canonical-ensemble corrections
relevant to the few-to-many-body crossover have been studied~\cite{Rizzi2018,SantAna2019}. 
A unified review of exact-solution
methods for strongly interacting trapped 1D quantum gases is given in
\cite{MinguzziVignolo2022}.

In this paper we derive and verify the canonical-ensemble (CE) scaling
of Tan's contact for the harmonically trapped Tonks--Girardeau gas at
fixed reduced temperature $\tau=T/T_F$ and large number of particles.
The contact admits a two-term large-$N$ expansion, with a leading
contribution of order $N^{5/2}$ and a subleading correction of order
$N^{3/2}$ governed by two universal $\tau$-dependent functions
$A(\tau)$ and $B(\tau)$; the precise scaling law is stated as
Eq.~\eqref{eq:scaling_law} at the start of
Sec.~\ref{sec:main_scaling}. The leading coefficient $A(\tau)$
coincides with the GCE result of
Ref.~\cite{VignoloMinguzzi2013}; we re-derive it from a contour-integral
representation of the canonical partition function and provide closed
asymptotic expansions in the Sommerfeld ($\tau\ll 1$) and Boltzmann
($\tau\gg 1$) limits. The subleading coefficient $B(\tau)$ is the focus
of the present work: we show that it admits a first-principles
saddle-point expression in terms of universal phase-space integrals of
the Fermi factor evaluated with a self-consistent scaled chemical
potential $\xi(\tau)$, that it is precisely the
ensemble-difference contribution between the canonical and
grand-canonical contacts at fixed mean particle number, and that it
admits universal closed asymptotic forms at both low and high
temperature. At low temperature $B(\tau)$ is linear in $\tau$ with a
negative slope, and at high temperature it tends to minus the leading
coefficient $A(\tau)$, the latter limit following from the Poissonian
particle-number statistics of the dilute Boltzmann regime. We
construct Pad\'e approximants for $A(\tau)$ and $B(\tau)$ that
interpolate between the two asymptotic regimes on a wide temperature
window, and verify the scaling law against canonical
contour-integration data. The result extends the few-body
canonical analysis of Ref.~\cite{SantAna2019} and
identifies the precise origin of the subleading $N^{3/2}$ term as a
finite-$N$ ensemble-correspondence effect. Read in light of recent
direct contact measurements in one-dimensional Lieb--Liniger gases
\cite{Huang2025}, the canonical scaling law provides a quantitative
target for trapped-geometry experiments at finite temperature.

The paper is organised as follows. Section~\ref{sec:preliminaries}
defines the model and establishes the canonical contour
representation of the contact. Section~\ref{sec:main_scaling}
contains the saddle-point analysis: the conventions and the
justification of the single-saddle reduction, the leading coefficient
$A(\tau)$ (Sec.~\ref{sec:CE_leading_scaling}), and the subleading
coefficient $B(\tau)$ (Sec.~\ref{sec:saddle_Btau}). The
Sommerfeld and virial derivations, including the boundary-layer
(``edge'') analysis, are collected in
Appendix~\ref{app:asymptotics}, and the numerical evaluation together
with the scaling verification is presented in
Sec.~\ref{sec:numerical_AB}. Section~\ref{sec:GCE} compares the
canonical and grand-canonical formulations, identifies $B(\tau)$ as
the ensemble correction, and provides its cumulant interpretation.
Section~\ref{sec:experimental} quantifies the experimental
implications of the scaling law for realistic particle numbers and
temperatures. We conclude in Sec.~\ref{sec:conclusions} with a
summary and outlook. Appendix~\ref{app:powercounting} collects the
power-counting arguments showing that no mechanism other than the
canonical saddle fluctuation contributes at order $N^{3/2}$, and
Appendix~\ref{app:numerical} documents the numerical procedure and
the Pad\'e approximants.

%======================================================================
\section{The model}
\label{sec:preliminaries}
%======================================================================

We consider $N$ identical bosons of mass $m$ on the line, confined by an
harmonic potential of frequency $\omega$ and interacting
through a contact potential of coupling strength $g>0$.
The Hamiltonian is
\begin{equation}
   H
  =\sum_{i=1}^{N} \left(-\frac{\hbar^{2}}{2m}\,\frac{\partial^{2}}{\partial x_{i}^{2}}
    +\frac{1}{2}m\omega^{2}x_{i}^{2}\right)
   +g \sum_{i<j}\delta(x_{i}-x_{j}).
  \label{eq:H_lieb_liniger_trap}
\end{equation}
Throughout this paper we work in harmonic-oscillator units, setting
$\hbar=m=\omega=k_{B}=1$, so that lengths are measured in units of the
oscillator length $a_{0}=\sqrt{\hbar/m\omega}$, energies in units of
$\hbar\omega$, and the single-particle eigenstates of the trap are the
familiar Hermite functions
\begin{equation}
  \phi_{n}(x)=\frac{e^{-x^{2}/2}H_{n}(x)}{\pi^{1/4}\sqrt{2^{n}n!}},
  \qquad
  \varepsilon_{n}=n+\frac{1}{2},
  \qquad n=0,1,2,\dots
  \label{eq:HO_eigenstates}
\end{equation}

We restrict our analysis to the Tonks--Girardeau (TG) limit
$g\to\infty$, in which the two-body interaction reduces to
the impenetrability constraint
$\Psi(x_{1},\dots,x_{N})=0$ whenever $x_{i}=x_{j}$ for any $i\neq j$.
In this limit the Bose--Fermi mapping of Girardeau~\cite{Girardeau1960}
expresses every bosonic eigenstate as a Slater determinant of
single-particle Hermite functions multiplied by an antisymmetrising
sign factor,
\begin{equation}
  \Psi^{(\mathrm b)}_{\alpha}(x_{1},\dots,x_{N})
  =\prod_{i<j}\mathrm{sgn}(x_{i}-x_{j})\;
   \Psi^{(\mathrm f)}_{\alpha}(x_{1},\dots,x_{N}),
  \label{eq:bose_fermi_map}
\end{equation}
where $\Psi^{(\mathrm f)}_{\alpha}$ is the noninteracting fermionic
Slater determinant labelled by the occupied set
$\alpha=\{n_{1},\dots,n_{N}\}$.
Since
$\left|\Psi^{(\mathrm b)}\right|^{2}=\left|\Psi^{(\mathrm f)}\right|^{2}$, all local one-body
observables of the TG gas (in particular the spatial density and the
energy) coincide with those of $N$ noninteracting spinless fermions in
the same trap.  Off-diagonal correlations, however, do not: the
Bose--Fermi map preserves probability density but not phase, so
quantities sensitive to the off-diagonal structure of the one-body
density matrix---most notably the momentum distribution---differ
remarkably between the two systems~\cite{Minguzzi2002,OlshaniiDunjko2003}.

The central object of this work is Tan's contact, defined through the
universal large-momentum tail of the single-particle momentum
distribution,
\begin{equation}
  \mathcal C
  =\lim_{k\to\infty}k^{4}\,n(k),
  \qquad
  n(k)=\frac{1}{2\pi} \int dx\,dx'\,e^{ik(x-x')}\,
  \rho^{(1)}(x,x'),
  \label{eq:contact_def}
\end{equation}
with $\rho^{(1)}$ the thermal one-body density matrix.  The 
contact admits the formulation as a one-dimensional integral
over a positive integrand $F_{N}(x)$\footnote{See Appendix \ref{app:derivation} for the detailed derivation.},
\begin{equation}
  \mathcal C_{N}
  =\frac{2}{\pi} \int_{-\infty}^{\infty} dx\,F_{N}(x),
  \qquad
  F_{N}(x)
  =\left\langle\rho(x)\,\kappa(x)-S(x)^{2}\right\rangle_{ N}^{(\mathrm{CE})},
  \label{eq:contact_FN}
\end{equation}
where the local objects $\rho$, $S$, $\kappa$ are diagonal and
near-diagonal values of the fermionic kernel and
$\langle\cdots\rangle_{N}^{(\mathrm{CE})}$ denotes the canonical thermal
average at fixed particle number $N$.  The representation
\eqref{eq:contact_FN} is the starting point for the saddle-point
analysis of the following sections. 

To compare different particle numbers and temperatures on a common
footing, we parametrise the temperature by the dimensionless ratio
\begin{equation}
  \tau:=\frac{T}{T_{F}},\qquad T_{F}=N\hbar\omega,
  \label{eq:tau_def}
\end{equation}
where $T_{F}$ is the Fermi temperature of the equivalent noninteracting
gas (the energy of the highest occupied harmonic level at $T=0$).  In
units with $\hbar=\omega=k_{B}=1$ this reads $T_{F}=N$.  At fixed
$\tau$, the inverse temperature scales as
$\beta:= 1/T=1/(\tau N)$, so that the thermal de Broglie wavelength
$\lambda_{T}=\sqrt{2\pi\beta}$ is parametrically smaller than both the
oscillator length and the Thomas-Fermi cloud size $R_{\mathrm{TF}}=\sqrt{2N}$
in the relevant scaling regime.  This separation of scales is what
makes the large-$N$ expansion at fixed $\tau$ a genuine semiclassical
limit and underlies the $N^{5/2}$ leading scaling of the contact
established in Sec.~\ref{sec:CE_leading_scaling}.

The two physically distinct regimes of $\tau$ correspond to the
familiar quantum statistics of a noninteracting Fermi gas: $\tau\ll1$
is the deeply degenerate regime in which the Fermi--Dirac distribution
is sharp and Sommerfeld-type expansions apply, while $\tau\gg 1$ is the
classical (Boltzmann) regime in which the occupation probabilities are
small and the partition function admits a virial expansion.  Both
limits are accessible analytically and serve as anchors for the
finite-$\tau$ analysis that follows.  In the intermediate regime
$\tau\sim 1$ we resort to numerical evaluation of the universal
scaling functions, supplemented by Pad\'e approximants that
interpolate between the two asymptotic forms.

\subsection{Kernel representation}
\begin{figure}[h!]
  \centering
  \includegraphics[width=0.85\textwidth]{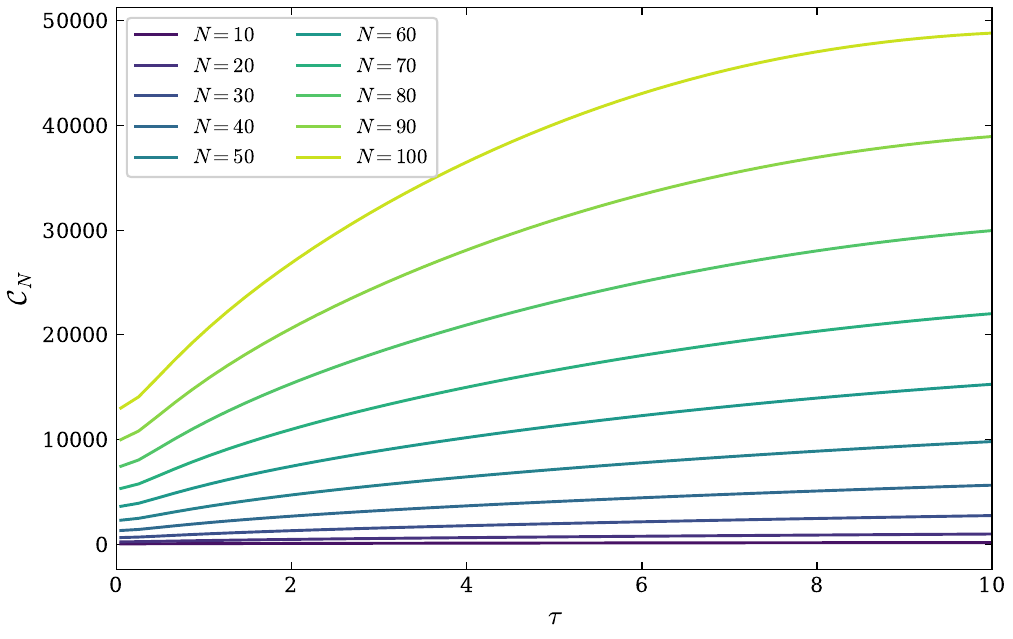}
  \caption{Canonical-ensemble contact \eqref{eq:contact_FN} as a function of the
    reduced temperature $\tau$, computed from the contour-integral
    representation \eqref{eq:FN_contour}. Curves correspond
    to particle numbers $N$ from 10 to 100 in steps of 10 (bottom to top). The data underlie the scaling-law
    verification of Eq.~\eqref{eq:scaling_law}.}
  \label{fig:contact_data}
\end{figure}

Let $\Xi(z)$ be the grand partition function of the trapped ideal Fermi gas,
\begin{equation}
\Xi(z)=\prod_{n\ge0}\left(1+z e^{-\beta\varepsilon_n}\right).
\label{eq:Xi_def_main}
\end{equation}
The canonical partition function $Z_N$ can be extracted through a coefficient 
analysis of $z^N$ in $\Xi(z)$,
admiting the contour representation
\begin{equation}
Z_N=\frac{1}{2\pi i}\oint_{\mathcal C}\frac{dz}{z^{N+1}}\,\Xi(z).
\label{eq:ZN_contour}
\end{equation}
The integrand $F_N(x)$ entering the contact \eqref{eq:contact_FN}
can be written as the corresponding contour average of a
grand-canonical kernel functional\footnote{See Appendix~\ref{app:kernel}
for the detailed derivation.}:
\begin{equation}
F_N(x)=\frac{1}{Z_N}\,\frac{1}{2\pi i}\oint_{\mathcal C}\frac{dz}{z^{N+1}}\,
\Xi(z)\,\left[\rho_z(x)\,\kappa_z(x)-S_z(x)^2\right],
\label{eq:FN_contour}
\end{equation}
where the kernel $K_z(x,y)$ and its local derivatives read
\begin{align}
K_z(x,y) &= \sum_{n\ge0} f_n(z)\,\phi_n(x)\phi_n(y),
\qquad
f_n(z)=\frac{z e^{-\beta\varepsilon_n}}{1+z e^{-\beta\varepsilon_n}},
\label{eq:kernel_def}
\\
\rho_z(x) &= K_z(x,x),\qquad
S_z(x)=\partial_y K_z(x,y)\big|_{y=x},\qquad
\kappa_z(x)=\partial_x\partial_y K_z(x,y)\big|_{y=x}.
\label{eq:local_defs}
\end{align}
Figure~\ref{fig:contact_data} shows the canonical contact $\mathcal C_N(\tau)$
evaluated from the contour-integral representation for a range of
particle numbers; the curves provide the raw data against which the
scaling law \eqref{eq:scaling_law} is verified in later sections.

%====================%====================%====================%====================
\section{Canonical contact scaling}
\label{sec:main_scaling}
% =====================================================================

In this section we derive the canonical-ensemble large-$N$ scaling law
\begin{equation}
\mathcal{C}_{N}(\tau) = A(\tau)\,N^{5/2} + B(\tau)\,N^{3/2},
\label{eq:scaling_law}
\end{equation}
through a contour-integral representation of the canonical partition
function followed by saddle-point reduction. The leading coefficient
$A(\tau)$ and its asymptotic limits are obtained in
Sec.~\ref{sec:CE_leading_scaling}; the subleading coefficient
$B(\tau)$, together with its explicit Sommerfeld and virial
expansions, is the subject of Sec.~\ref{sec:saddle_Btau}.

% =====================================================================
\subsection{Leading term}
\label{sec:CE_leading_scaling}
In order to extract the large-$N$ behaviour, we evaluate both contour
integrals \eqref{eq:ZN_contour} and \eqref{eq:FN_contour} by steepest
descent. We first fix the conventions of the saddle-point analysis. The points collected here also underpin the subleading analysis of
Sec.~\ref{sec:saddle_Btau}.

It is convenient to introduce the variable $t$ through $z=e^{t}$,
understood as a local complex coordinate in a neighbourhood of
the positive real axis. 
Under this substitution the measure of both contour integrals
transforms as
\begin{equation}
\frac{dz}{z^{N+1}} = e^{-Nt}\,dt ,
\label{eq:measure_exact}
\end{equation}
so that, defining
\begin{equation}
\psi(t):=\log\Xi(e^{t}),
\qquad
\Phi(t):=\psi(t)-Nt,
\label{eq:Phi_def}
\end{equation}
both integrals take the form $\oint dt\, e^{\Phi(t)}\,[\cdots]$. The
offset between the exponent $-(N{+}1)\log z$ of the original
integrand and the $-Nt$ appearing in $\Phi$ is absorbed by
the Jacobian $dz=e^{t}dt$: in the $t$-representation there is no
residual $\mathcal O(1/N)$ ambiguity in the definition of the saddle.
Throughout this paper the saddle point $t_\star$ is defined as the
solution of $\Phi'(t_\star)=0$, i.e.\ of the number equation
\begin{equation}
N=\sum_{n\ge0}\frac{z_\star e^{-\beta\varepsilon_n}}{1+z_\star e^{-\beta\varepsilon_n}}
=\sum_{n\ge0} f_n(z_\star),
\qquad z_\star=e^{t_\star},
\label{eq:saddle_number}
\end{equation}
with exactly $N$ (not $N{+}1$) on the left-hand side. Since
$\psi'(t)=\sum_n f_n(e^{t})$ increases strictly and continuously from
$0$ to $\infty$ along the real axis
[$\psi''(t)=\sum_n f_n(1-f_n)>0$], the real saddle $t_\star$ exists
and is unique. It is convenient to parametrize $z_\star=e^{\beta\mu}$, so that
\begin{equation}
f_n(z_\star)=\frac{1}{e^{\beta(\varepsilon_n-\mu)}+1},
\label{eq:FD_at_saddle}
\end{equation}
where $\mu=\mu(N,T)$ is the chemical potential.

We next specify the contour and the expansion point, which is what
makes the passage from the contour representations
\eqref{eq:ZN_contour} and \eqref{eq:FN_contour} to a single-saddle
Gaussian expansion transparent.

(i) Choice of contour and dominance of the real saddle. The
grand partition function $\Xi(z)$ of the trapped ideal Fermi gas is an analytic function of $z$ throughout the complex plane, and so is the combination
$\Xi(z)\left[\rho_z(x)\kappa_z(x)-S_z(x)^{2}\right]$ appearing under
the contour in \eqref{eq:FN_contour}: the simple poles of the
occupation factors $f_n(z)$ at $z=-e^{\beta\varepsilon_n}$ on the
negative real axis are cancelled by the zeros of $\Xi(z)$ at the same
points. The only singularity of either integrand is the pole at
$z=0$, so the contour $\mathcal C$ may be deformed freely; we take it
to be the circle through the saddle,
\begin{equation}
z=z_\star e^{i\theta}
\quad\Longleftrightarrow\quad
t=t_\star+i\theta,
\qquad \theta\in(-\pi,\pi] .
\label{eq:contour_circle}
\end{equation}
On this circle the modulus of the dominant factor is controlled by
\begin{equation}
\mathrm{Re}\,\psi(t_\star+i\theta)
=\frac12\sum_{n\ge0}\log\left[1+2z_\star e^{-\beta\varepsilon_n}\cos\theta
+z_\star^{2}e^{-2\beta\varepsilon_n}\right],
\label{eq:modulus_on_circle}
\end{equation}
in which every summand is strictly decreasing in $|\theta|$ on
$[0,\pi]$. The modulus of the integrand therefore attains its unique
maximum at $\theta=0$ and decays monotonically toward
$\theta=\pm\pi$: the circle \eqref{eq:contour_circle} is a globally
admissible steepest-descent-type contour, the positive-real saddle
$z_\star$ strictly dominates it, and no other saddle of the
complexified exponent lies on, or contributes to, the deformed
contour. Near $\theta=0$,
$\Phi(t_\star+i\theta)\simeq\Phi(t_\star)-\frac12\,\Phi''(t_\star)\,\theta^{2}$
with $\Phi''(t_\star)=\sum_n f_n(1-f_n)>0$: the steepest-descent
direction through $t_\star$ is the angular direction
$t=t_\star+i\theta$, along which the integrand decays as a Gaussian
of variance $1/\Phi''(t_\star)=\mathcal O(1/N)$, and contributions
from $|\theta|\gtrsim N^{-1/2}$ are exponentially suppressed. All
saddle-point expansions in this paper are Gaussian expansions in
$\theta$ along this contour.

(ii) Common expansion point for numerator and denominator. The
numerator \eqref{eq:FN_contour} contains the additional $t$-dependent
factor $G(e^{t};x):=\rho_z(x)\kappa_z(x)-S_z(x)^{2}$, which displaces
its effective saddle from $t_\star$ by
$\delta t=-\left.\partial_t\log G\right|_{t_\star}/\Phi''(t_\star)
=\mathcal O(1/N)$. We do not recentre the numerator at its own
effective saddle: both integrals are expanded about the common point
$t_\star$ defined by \eqref{eq:saddle_number}, and the displacement
$\delta t$ is retained at the working order through the
first-derivative term of the Laplace ratio expansion,
Eq.~\eqref{eq:ratio_expansion_t} below. This bookkeeping is
irrelevant for the leading $N^{5/2}$ term but matters at order
$N^{3/2}$. See Sec.~\ref{sec:saddle_Btau} and
Appendix~\ref{app:powercounting}.

Since both numerator and denominator of \eqref{eq:FN_contour} are
dominated by the same saddle $z_\star$, the leading contribution to
their ratio is obtained by evaluating the kernel functional at
$z_\star$:
\begin{equation}
F_N(x)\simeq \rho_{z_\star}(x)\,\kappa_{z_\star}(x)-S_{z_\star}(x)^2+\cdots,
\label{eq:FN_saddle_reduction}
\end{equation}
where the displayed equality is asymptotic (the omitted terms are
the Gaussian saddle-fluctuation corrections, which we treat
systematically in Sec.~\ref{sec:saddle_Btau}).

% ---------------------------------------------------------------------

We now evaluate $\rho_{z_\star}(x)$, $\kappa_{z_\star}(x)$ and $S_{z_\star}(x)$
in the large-$N$, semiclassical limit at fixed $\tau$.
In the Wigner, or phase-space, approximation, the kernel becomes local in $(x,p)$
with Fermi occupation
\begin{equation}
f(p,x)=\frac{1}{e^{\beta\left(\frac{p^2}{2}+V(x)-\mu\right)}+1},
\qquad
V(x)=\frac{x^2}{2}.
\label{eq:phase_space_f}
\end{equation}
Then the local objects defined in \eqref{eq:local_defs} reduce to the 
moments
\begin{align}
& \rho_{z_\star}(x)
\simeq \int_{-\infty}^{\infty}\frac{dp}{2\pi}\, f(p,x),
\label{eq:rho_phase_space}
\\
& S_{z_\star}(x)
\simeq \int_{-\infty}^{\infty}\frac{dp}{2\pi}\, (ip)\, f(p,x)=0,
\label{eq:S_zero}
\\
& \kappa_{z_\star}(x)
\simeq \int_{-\infty}^{\infty}\frac{dp}{2\pi}\, p^2\, f(p,x).
\label{eq:kappa_phase_space}
\end{align}
The vanishing of $S(x)$ follows from the parity $p\mapsto -p$ of $f(p,x)$.

Firstly, note that \eqref{eq:phase_space_f} replaces the exact Wigner symbol of the
operator $f_z(\hat h)$ by the classical Fermi factor. The omitted
corrections are of relative order $1/N^{2}$
at fixed $\tau$. Second, \eqref{eq:S_zero} is a leading-order
phase-space statement: the symmetry
$K_z(x,y)=K_z(y,x)$ gives $S_z(x)=\frac12\,\rho_z'(x)$,
which is $\mathcal O(1)$ in the scaling regime, so
that $S_z^{2}$ contributes to the contact only at order $N^{1/2}$.
Third, the same $1/N^{2}$ counting applies to the centre-gradient
corrections to \eqref{eq:rho_phase_space} and
\eqref{eq:kappa_phase_space}. None of these corrections affects the
$N^{5/2}$ or the $N^{3/2}$ term. The systematic power counting is
given in Appendix~\ref{app:powercounting}.

Using \eqref{eq:FN_saddle_reduction} and \eqref{eq:S_zero}, we obtain
\begin{equation}
F_N(x)\simeq \rho(x)\,\kappa(x).
\label{eq:FN_rho_kappa}
\end{equation}

% ---------------------------------------------------------------------

We now show that \eqref{eq:FN_rho_kappa} implies $\mathcal C_N(\tau)\propto N^{5/2}$.
At fixed $\tau$ we have $\beta\sim 1/(\tau N)$, and the saddle chemical
potential $\mu(N,\tau)$ is fixed by the particle-number constraint
\eqref{eq:saddle_number}. Anticipating the scaling
$\mu = \xi(\tau)\,N$ that will be made explicit shortly,
we introduce the standard large-$N$ scaling of coordinates and momenta,
\begin{equation}
x=\sqrt{2N}\,u,\qquad p=\sqrt{2N}\,q.
\label{eq:scaling_xp}
\end{equation}
Then
\begin{equation}
\frac{p^2}{2}+V(x)-\mu
= N(q^2+u^2-\xi),
\label{eq:energy_scaling}
\end{equation}
so that, using $\beta=1/(\tau N)$,
\begin{equation}
\beta\left(\frac{p^2}{2}+V(x)-\mu\right)
=\frac{q^2+u^2-\xi}{\tau}.
\label{eq:beta_energy_scaling}
\end{equation}
Therefore the occupation \eqref{eq:phase_space_f} becomes $N$-independent at
leading order,
\begin{equation}
f(p,x)\ \longrightarrow\ f_\tau(q,u)
:= \frac{1}{e^{(q^2+u^2-\xi)/\tau}+1},
\qquad \xi=\xi(\tau).
\label{eq:f_tau}
\end{equation}

The scaled chemical potential $\xi(\tau)$ is determined self-consistently
by the large-$N$ limit of the number constraint
\eqref{eq:saddle_number}. Replacing the discrete sum over
single-particle levels by an integral against the 1D harmonic
oscillator density of states, and using
$\beta = 1/(\tau N)$, $\mu = \xi N$, $\varepsilon_n = NE$ recasts
\eqref{eq:saddle_number} as the transcendental equation
\begin{equation}
\tau\,\log\left(1+e^{\xi/\tau}\right) = 1\,,
\qquad \xi = \xi(\tau).
\label{eq:xi_eq}
\end{equation}
Equation \eqref{eq:xi_eq} admits the asymptotic behaviour
$\xi(\tau\ll 1) = 1 - \tau\,e^{-1/\tau}+\mathcal O(e^{-2/\tau})$ at
low temperature and $\xi(\tau\gg 1)\sim -\tau\log\tau$ at high
temperature. The zero-temperature limit $\xi(0)=1$ follows directly
from \eqref{eq:xi_eq}: as $\tau\to 0^+$, $\log(1+e^{\xi/\tau})\to
\max(\xi/\tau,0)$. Thus the
chemical potential coincides with $N$ at $T=0$; the corrections at
small $\tau$ are exponentially suppressed and therefore invisible in
the Sommerfeld series, while at $\tau\gtrsim 1$ they grow
polynomially and must be retained. For all numerical results below we use the
self-consistent $\xi(\tau)$ from \eqref{eq:xi_eq}: the simpler choice
$\xi = 1$ correctly reproduces the low-$\tau$ Sommerfeld expansion
but introduces uncontrolled $\mathcal O(\tau\log\tau)$ errors at
$\tau\gtrsim 1$ that propagate to the universal functions $A(\tau)$
and $B(\tau)$.

Let us begin by defining the dimensionless moments
\begin{equation}
I_0(u;\tau):= \int_{-\infty}^{\infty}dq\, f_\tau(q,u),
\qquad
I_2(u;\tau):= \int_{-\infty}^{\infty}dq\, q^2\, f_\tau(q,u).
\label{eq:I0_I2_def}
\end{equation}
Using $dp=\sqrt{2N}\,dq$ and \eqref{eq:rho_phase_space}--\eqref{eq:kappa_phase_space}
gives
\begin{align}
\rho(x) &\simeq \int\frac{dp}{2\pi} f(p,x)
= \frac{\sqrt{2N}}{2\pi}\, I_0(u;\tau),
\label{eq:rho_scaling}
\\
\kappa(x) &\simeq \int\frac{dp}{2\pi} p^2 f(p,x)
= \frac{(2N)^{3/2}}{2\pi}\, I_2(u;\tau).
\label{eq:kappa_scaling}
\end{align}
Inserting \eqref{eq:rho_scaling} and \eqref{eq:kappa_scaling} into
\eqref{eq:FN_rho_kappa} yields
\begin{equation}
F_N(x)\simeq \rho(x)\kappa(x)
= \frac{(2N)^2}{4\pi^2}\, I_0(u;\tau)\,I_2(u;\tau)
= \frac{N^2}{\pi^2}\, I_0(u;\tau)\,I_2(u;\tau).
\label{eq:FN_scaling}
\end{equation}

Finally, using the definition of the contact \eqref{eq:contact_def}, we obtain the large-$N$ limit  leading scaling term 
\begin{align}
\mathcal C_N(\tau)
&\simeq \frac{2}{\pi}\int_{-\infty}^{\infty}dx\,
\frac{N^2}{\pi^2}\, I_0(u;\tau)\,I_2(u;\tau)
\nonumber\\
&= \frac{2}{\pi}\,\sqrt{2N}\,\frac{N^2}{\pi^2}
\int_{-\infty}^{\infty}du\, I_0(u;\tau)\,I_2(u;\tau)
\nonumber\\
&= A(\tau)\,N^{5/2},
\label{eq:CN_leading}
\end{align}
where we defined the universal scaling function $A(\tau)$ as 
\begin{equation}
A(\tau):= \frac{2\sqrt{2}}{\pi^3}\int_{-\infty}^{\infty}du\, I_0(u;\tau)\,I_2(u;\tau),
\label{eq:A_tau_def}
\end{equation}
with $I_0,I_2$ given by \eqref{eq:I0_I2_def} and $f_\tau$ by \eqref{eq:f_tau}.

\subsubsection{Low temperature}
\label{sec:A_lowtau}
In the zero-temperature limit, $f_\tau(q,u)\to \Theta(1-u^2-q^2)$.
Then for $|u|<1$,
\begin{equation}
I_0(u;0)=2\sqrt{1-u^2},
\qquad
I_2(u;0)=\frac{2}{3}(1-u^2)^{3/2},
\qquad
I_0 I_2=\frac{4}{3}(1-u^2)^2,
\end{equation}
and $I_0=I_2=0$ for $|u|>1$. Therefore
\begin{equation}
\int_{-\infty}^{\infty}du\, I_0(u;0)\,I_2(u;0)
=\frac{4}{3}\int_{-1}^{1}du\, \left(1-u^2\right)^2
=\frac{64}{45},
\end{equation}
which gives
\begin{equation}
A(0)
=\frac{128\sqrt2}{45\pi^3},
\end{equation}
reproducing the known $T=0$ coefficient.

At low finite temperature, the corrections to $\xi(\tau)$ are
exponentially small\footnote{Recall
$\xi(\tau\ll 1)=1-\tau e^{-1/\tau}+\mathcal O(e^{-2/\tau})$ from
\eqref{eq:xi_eq}.}, and a standard Sommerfeld expansion of the moments
$I_0$ and $I_2$ about the local Fermi level, carried out in
Appendix~\ref{app:asymptotics_A_low}, produces the
simple bulk result
$I_0I_2=\frac43(1-u^2)^2+\frac{\pi^2}{9}\tau^2$ for $|u|<1$, with a
$\tau^{2}$ term independent of $u$. The turning-point regions
$|u|\simeq1$, of width $\Delta u\sim\tau$, contribute only at
$\mathcal O(\tau^{3})$ and do not affect the $\tau^{2}$ coefficient.
Integrating over $u$ gives
\begin{equation}
A(\tau)
=
\frac{2\sqrt2}{\pi^3}
\left(
\frac{64}{45}+\frac{2\pi^2}{9}\tau^2
\right)
=
\underbrace{\frac{128\sqrt2}{45\pi^3}}_{A(0)}
+\underbrace{\frac{4\sqrt2}{9\pi}}_{a_2}\,\tau^2,
\qquad (\tau\ll 1).
\label{eq:A_tau_lowtau}
\end{equation}

It is sometimes convenient to write the correction in relative form:
\begin{equation}
A(\tau)=A(0)\left[1+\frac{a_2}{A(0)}\tau^2+\cdots\right]
=
A(0)\left[1+\frac{5\pi^2}{32}\tau^2 \right],
\qquad
A(0)=\frac{128\sqrt2}{45\pi^3}.
\label{eq:A_tau_relative}
\end{equation}

% =====================================================================
\subsubsection{High temperature}
\label{sec:A_hightau}
% =====================================================================
For $\tau\gg 1$ the saddle fugacity $z:=e^{\xi/\tau}$ is small: the
LDA number constraint \eqref{eq:xi_eq} reads $\tau\log(1+z)=1$, hence
$z=e^{1/\tau}-1=1/\tau+\mathcal O(1/\tau^{2})\ll1$, and the gas is in
the Boltzmann regime. Expanding the Fermi factor in the virial
(small-$z$) series, the moments \eqref{eq:I0_I2_def} reduce to
elementary Gaussian integrals; carrying the expansion through second
order in $z$ and eliminating $z$ in favour of $\tau$ via the number
constraint (Appendix~\ref{app:asymptotics_A_high}) yields
\begin{equation}
A(\tau)=\frac{\sqrt{\tau}}{\pi^{3/2}}
\left(
1+\frac{2-\sqrt3}{2\tau}
+\mathcal O \left(\tau^{-2}\right)
\right),
\qquad (\tau\gg 1).
\label{eq:A_hightau_final}
\end{equation}

%======================================================
\subsection{Subleading term}
\label{sec:saddle_Btau}

To extract the subleading $N^{3/2}$ correction we go beyond the
leading saddle evaluation \eqref{eq:FN_saddle_reduction} of the
canonical contour representation \eqref{eq:FN_contour} and include
the Gaussian fluctuations around the saddle. In the
$t$-representation of Eqs.~\eqref{eq:measure_exact} and
\eqref{eq:Phi_def}, both contour integrals are Laplace-type
integrals along the steepest-descent circle
\eqref{eq:contour_circle},\footnote{The symbol
$\Phi$ also appears as the universal edge functions
$\Phi_0(y),\Phi_2(y)$ of \eqref{eq:Phi0Phi2_def}; the saddle action
$\Phi(t)$ used here, a function of $t=\log z$, is unrelated.
Subscripts $\Phi_k=\Phi^{(k)}(t_\star)$ in this subsection denote
derivatives of the saddle action, while the subscripts on
$\Phi_0,\Phi_2$ denote moment orders.}
with the saddle condition $\Phi'(t_\star)=0$ being the number constraint
\eqref{eq:saddle_number} and the Fermi factors \eqref{eq:FD_at_saddle} evaluated at
$z_\star=e^{t_\star}=e^{\beta\mu}$. Expanding $\Phi$ and
$\widetilde G(t;x):=G(e^t;x)$ around the common point $t_\star$, the standard Laplace
ratio expansion yields
\begin{equation}
F_N(x)=\widetilde G_\star(x)
-\frac{1}{2}\frac{\widetilde G_2(x)}{\Phi_2}
+\frac{1}{2}\frac{\Phi_3}{\Phi_2^2}\,\widetilde G_1(x)
+R_N(x),
\label{eq:ratio_expansion_t}
\end{equation}
with $\Phi_k=\Phi^{(k)}(t_\star)$ and
$\widetilde G_k=\partial_t^k\widetilde G\big|_{t_\star}$. The
remainder $ R_N(x)$ is suppressed by an additional factor of
$1/\Phi_2 \sim 1/N$ relative to the displayed corrections, i.e.\
$ R_N = \mathcal O(\widetilde G_\star/N^2)$ in absolute terms or
$\mathcal O(N^{-2})$ relative to the leading $\widetilde G_\star$.
As anticipated in Sec.~\ref{sec:CE_leading_scaling}, the
$\mathcal O(1/N)$ displacement of the numerator's effective saddle
relative to $t_\star$ is not neglected in
\eqref{eq:ratio_expansion_t}: it is carried by the term
$\propto\widetilde G_1$, as one may verify by recentring the
numerator at its displaced saddle
$t_\star+\delta t$ and re-expanding, which reproduces
\eqref{eq:ratio_expansion_t} term by term at this order
(Appendix~\ref{app:powercounting}).
Since $\psi(t)$
is the cumulant generating function of the grand-canonical particle
number, $\Phi_2$ and $\Phi_3$ are themselves particle-number
cumulants:
\begin{equation}
\Phi_2 = \kappa_2 = \sum_{n\ge0}f_n(1-f_n),\qquad
\Phi_3 = \kappa_3 = \sum_{n\ge0}f_n(1-f_n)(1-2f_n),
\label{eq:cumulants_discrete}
\end{equation}
both of which scale as $\mathcal O(N)$ at fixed $\tau$, making the
two displayed correction terms in \eqref{eq:ratio_expansion_t}
of relative order $1/N$ compared to $\widetilde G_\star$
(absolute order $\widetilde G_\star/N$). 

In the fixed-$\tau$ scaling limit, parity again kills $S(x)$ and the
leading saddle functional reduces, via \eqref{eq:rho_scaling} and
\eqref{eq:kappa_scaling}, to
\begin{equation}
\widetilde G_\star(x)\simeq \rho(x)\kappa(x)
\simeq \frac{N^2}{\pi^2}\,g_0(u;\tau),
\qquad
g_0(u;\tau):= I_0(u;\tau)\,I_2(u;\tau),
\label{eq:Gstar_scaling_g0}
\end{equation}
with $I_0,I_2$ given by \eqref{eq:I0_I2_def}\footnote{With the leading scaling
$\widetilde G_\star \sim N^2$ established in 
\eqref{eq:Gstar_scaling_g0}, the displayed corrections are
$\mathcal O(N)$ in absolute terms, and the remainder
$ R_N = \mathcal O(1)$.}. The dependence of
the Fermi factor \eqref{eq:f_tau} on $t=\beta\mu$ enters only through
$f_\tau$, with
\begin{equation}
\partial_t f_\tau = f_\tau(1-f_\tau),\qquad
\partial_t^2 f_\tau = f_\tau(1-f_\tau)(1-2f_\tau),
\label{eq:t_derivatives_f}
\end{equation}
so differentiation under the $q$-integral introduces the local
integrals
\begin{align}
J_0(u;\tau)&:=\int_{-\infty}^{\infty}dq\, f_\tau(1-f_\tau),
&
J_2(u;\tau)&:=\int_{-\infty}^{\infty}dq\, q^2 f_\tau(1-f_\tau),
\label{eq:J0_J2}
\\
K_0(u;\tau)&:=\int_{-\infty}^{\infty}dq\, f_\tau(1-f_\tau)(1-2f_\tau),
&
K_2(u;\tau)&:=\int_{-\infty}^{\infty}dq\, q^2 f_\tau(1-f_\tau)(1-2f_\tau).
\label{eq:K0_K2}
\end{align}
Using $\partial_t I_\alpha=J_\alpha$, $\partial_t^2 I_\alpha=K_\alpha$
and $g_0=I_0 I_2$,
\begin{align}
\partial_t g_0 &= J_0 I_2 + I_0 J_2,
\label{eq:g0_t}
\\
\partial_t^2 g_0 &= K_0 I_2 + 2J_0J_2 + I_0K_2.
\label{eq:g0_tt}
\end{align}

Importantly, a phase-space replacement of the discrete sums in
\eqref{eq:cumulants_discrete} converts the cumulants into global
$u$-integrals,
\begin{equation}
\kappa_2 \simeq \frac{N}{\pi}\,V_2(\tau),
\qquad
V_2(\tau):=\int_{-\infty}^{\infty}du\,J_0(u;\tau),
\label{eq:kappa2_scaling}
\end{equation}
\begin{equation}
\kappa_3 \simeq \frac{N}{\pi}\,V_3(\tau),
\qquad
V_3(\tau):=\int_{-\infty}^{\infty}du\,K_0(u;\tau).
\label{eq:kappa3_scaling}
\end{equation}
Inserting \eqref{eq:Gstar_scaling_g0}--\eqref{eq:kappa3_scaling} into
\eqref{eq:ratio_expansion_t} gives us
\begin{equation}
F_N(x)=\frac{N^2}{\pi^2}\,g_0(u;\tau)
-\frac{N}{\pi}\,\mathcal H(u;\tau),
\label{eq:FN_saddle}
\end{equation}
with
\begin{equation}
\mathcal H(u;\tau):=
\frac{1}{2}\left[
\frac{\partial_t^2 g_0(u;\tau)}{V_2(\tau)}
-\frac{V_3(\tau)}{V_2(\tau)^2}\,\partial_t g_0(u;\tau)
\right].
\label{eq:H_def}
\end{equation}
Therefore, inserting \eqref{eq:FN_saddle} into \eqref{eq:contact_FN} and integrating (with 
$dx=\sqrt{2N}\,du$) then yields the canonical scaling law \eqref{eq:scaling_law} 
with $A(\tau)$ as in \eqref{eq:A_tau_def} and the new universal
subleading function
\begin{equation}
B(\tau)
=-\frac{2\sqrt{2}}{\pi^2}\int_{-\infty}^{\infty}du\,\mathcal H(u;\tau).
\label{eq:Bsad_master}
\end{equation}
Substituting \eqref{eq:g0_t}--\eqref{eq:H_def} into
\eqref{eq:Bsad_master}, we obtain the explicit form
\begin{equation}
B(\tau)
=
\frac{\sqrt2}{\pi^2}\int_{-\infty}^{\infty}du\,
\left[
\frac{V_3(\tau)}{V_2(\tau)^2}\,(J_0 I_2+I_0 J_2)
-\frac{K_0 I_2+2J_0J_2+I_0K_2}{V_2(\tau)}
\right],
\label{eq:Bsad_oneline}
\end{equation}
which expresses the subleading coefficient entirely in terms of
universal phase-space integrals of the Fermi factor, valid for all
$\tau\ge 0$.

The result \eqref{eq:Bsad_oneline} is complete at order $N^{3/2}$
only if no mechanism other than the Gaussian saddle fluctuation
contributes at that order, and this requires an argument: the
leading contact density scales locally as $N^{2}$ and is integrated over a
region of size $\mathcal O(\sqrt N)$, so any correction of
relative order $1/N$ to the local integrand would feed the
$N^{3/2}$ coefficient. In Appendix~\ref{app:powercounting} we
enumerate the possible sources (the displacement of the numerator
saddle, the discreteness corrections to the number equation, the
corrections to the phase-space Fermi factor, the centre-gradient corrections to the kernel quantities
$\rho_z,S_z,\kappa_z$, and the finite-$N$ corrections to the
grand-canonical reference of Sec.~\ref{sec:GCE}) and show that
each of them enters only at relative order $1/N^{2}$, i.e.\ at
order $N^{1/2}$ in the contact. A special role is played by the
midpoint structure $\varepsilon_n=n+\frac12$ of the oscillator
spectrum, which eliminates the $\mathcal O(1)$ Euler--Maclaurin
boundary term in the number equation that would otherwise shift the
chemical potential. The coefficient
$B(\tau)$ in \eqref{eq:Bsad_oneline} is therefore the complete
$N^{3/2}$ coefficient of the canonical contact.

For the numerical verification of the scaling law
\eqref{eq:scaling_law} carried out throughout the rest of the
paper, it is convenient to introduce the dimensionless ratio
\begin{equation}
  \mathcal R_N(\tau)
  := \frac{\mathcal C_N(\tau)}{A(\tau)\,N^{5/2}+B(\tau)\,N^{3/2}},
  \label{eq:RN_def}
\end{equation}
which satisfies $\mathcal R_N(\tau)\to 1$ as $N\to\infty$ at fixed
$\tau$, with relative deviation $\mathcal O(N^{-1})$ controlled by the
next-to-subleading $N^{1/2}$ correction. We now evaluate
\eqref{eq:Bsad_oneline} in the asymptotic limits $\tau\ll 1$ and
$\tau\gg 1$.

% =====================================================================
\subsubsection{Low temperature}
\label{sec:Bsad_lowtau}
% =====================================================================

In the degenerate regime the subleading coefficient is governed by
the same bulk-plus-edge structure as the Sommerfeld analysis of
$A(\tau)$, with one essential refinement: the derivatives of the
Fermi factor generate negative powers of $a:=1-u^{2}$ that render the
bulk expansion nonuniform near the turning points, so the two edge
layers $|1-u^{2}|\sim\tau$ must be resolved separately. We summarise the method and the results here.
The detailed calculations are given in
Appendices~\ref{app:asymptotics_B_low} and
\ref{app:asymptotics_B_edge}.

In the bulk $|u|<1$, Sommerfeld expansion of the six phase-space
integrals \eqref{eq:I0_I2_def}, \eqref{eq:J0_J2} and \eqref{eq:K0_K2}
gives (Appendix~\ref{app:asymptotics_B_low})
\begin{equation}
\partial_t^2 g_0(u;\tau)=\frac{8}{3}\,\tau^2+\mathcal O(\tau^4),
\label{eq:dtg2_bulk_main}
\end{equation}
independent of $u$, while the global cumulant functions
obey the exact phase-space identities
\begin{equation}
V_2(\tau)=\pi\tau\,f(-\xi/\tau),
\qquad
V_3(\tau)=\pi\tau\,f(-\xi/\tau)\left[1-f(-\xi/\tau)\right],
\qquad f(y)=\frac{1}{e^{y}+1},
\label{eq:V2V3_exact_main}
\end{equation}
so that $V_2=\pi\tau\,[1+\mathcal O(e^{-1/\tau})]$ while
$V_3\sim\pi\tau\,e^{-1/\tau}$ is nonperturbatively small: the entire
$V_3$-term of \eqref{eq:H_def} contributes only at
$\mathcal O(e^{-1/\tau})$, and the algebraic expansion of $B(\tau)$
is controlled solely by $\partial_t^2 g_0/V_2$. The bulk
contribution to \eqref{eq:Bsad_master} is therefore linear in
$\tau$.

The turning-point layers require separate treatment. In the edge
variable $y=(1-u^2)/\tau$, with momenta rescaled as $q=\sqrt\tau\,p$,
the Fermi factor collapses to the universal profile
$f_y(p)=(e^{p^2-y}+1)^{-1}$, whose moments
\begin{equation}
\Phi_0(y):=\int_{-\infty}^{\infty}dp\, f_y(p),
\qquad
\Phi_2(y):=\int_{-\infty}^{\infty}dp\, p^2\, f_y(p),
\label{eq:Phi0Phi2_def}
\end{equation}
build a single universal edge function
$\mathcal P(y):=\Phi_0(y)\Phi_2(y)$, in terms of which the
combination entering $\partial_t^2 g_0$ collapses to a total second
derivative, $K_0I_2+2J_0J_2+I_0K_2=\tau^2\,\mathcal P''(y)$. Each of
the two layers consists of an inner part ($y>0$, inside the cloud)
and an outer part ($y<0$, the classically forbidden side), and a
correspondence analysis shows that their respective
contributions to the $\tau^{2}$ coefficient of $B(\tau)$,
$+\frac{\sqrt2}{\pi^3}\,\mathcal P'(0)$ and
$-\frac{\sqrt2}{\pi^3}\,\mathcal P'(0)$, cancel 
(Appendix~\ref{app:asymptotics_B_edge}). The edge layers therefore
leave no trace at order $\tau^{2}$, and the linear law extends
through that order,
\begin{equation}
B(\tau)
=
-\frac{16\sqrt2}{3\pi^3}\,\tau
+\mathcal O\left(\tau^3\right)+\mathcal O\left(e^{-1/\tau}\right),
\qquad (\tau\ll1).
\label{eq:Bsad_lowtau_final_closed}
\end{equation}

The expansion \eqref{eq:Bsad_lowtau_final_closed} (and, more
generally, the entire $B(\tau)$ analysis) is performed at fixed
$\tau>0$ with $N\to\infty$, and the limits $N\to\infty$ and
$\tau\to0$ do not commute. The bulk-plus-edge treatment
requires the thermal width of the Fermi surface to exceed its
quantum width: the thermal edge layer has width $\sim\tau$ in the
scaled variable $u$, while at strictly zero temperature the density
profile terminates over the Airy scale $\sim N^{-2/3}$. The present
low-$\tau$ analysis therefore holds in the window
\begin{equation}
\tau \gg N^{-2/3},
\label{eq:validity_window}
\end{equation}
i.e.\ $\tau\gtrsim0.2$ at $N=10$ and $\tau\gtrsim0.05$ at $N=100$.
Below this crossover the zero-temperature finite-$N$ hierarchy takes
over, with edge corrections scaling as $N^{3/4}$ and $N^{1/4}$
\cite{VignoloMinguzzi2013,Rizzi2018}. Since $B(\tau)\to0$ linearly,
the $N^{3/2}$ term derived here becomes subdominant to those
zero-temperature edge terms in the strict $\tau\to0$ limit at fixed
$N$. The same window \eqref{eq:validity_window} delimits the
Wigner--Kirkwood power counting of
Appendix~\ref{app:powercounting}, whose gradient corrections grow to
$\mathcal O(1)$ at $\tau\sim N^{-2/3}$. Experimental
consequences of this window are discussed in
Sec.~\ref{sec:experimental}.

\begin{figure}[h!]
  \centering
  \includegraphics[width=\textwidth]{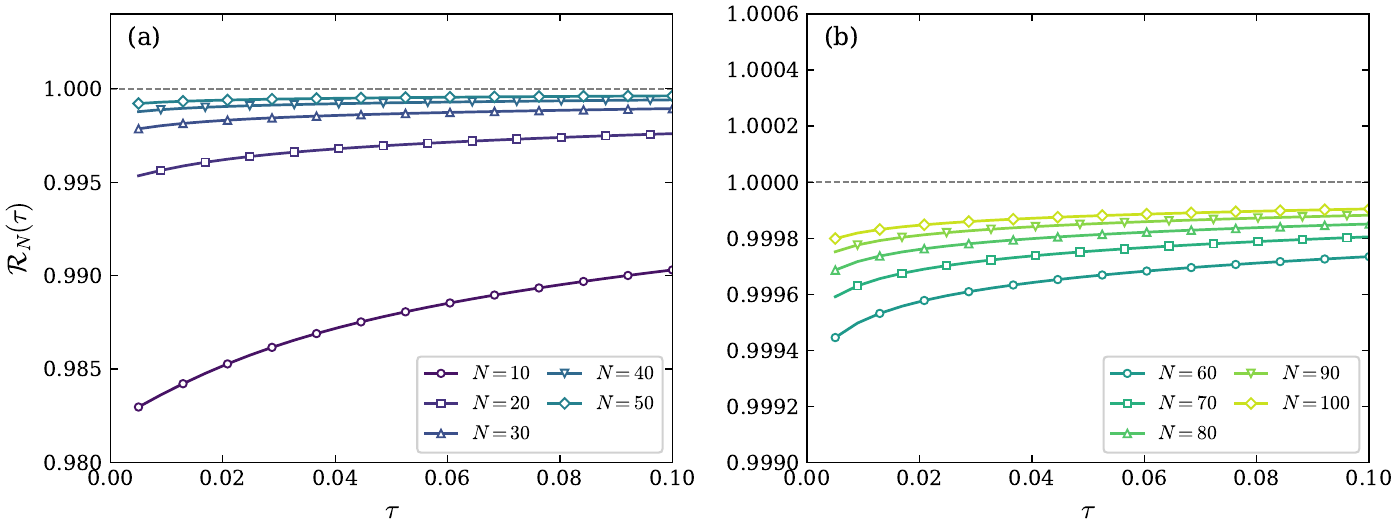}
  \caption{Low-$\tau$ verification of the canonical scaling law
    \eqref{eq:scaling_law}. The
    ratio $\mathcal R_N$ defined in \eqref{eq:RN_def}, with
    $A(\tau)$ and $B(\tau)$ obtained from the universal integral
    representations \eqref{eq:A_tau_def} and \eqref{eq:Bsad_oneline}
    using the self-consistent scaled chemical potential $\xi(\tau)$
    from \eqref{eq:xi_eq}, on $\tau\in[0.005,0.1]$:
    (a) $N=10,\dots,50$; (b) $N=60,\dots,100$ [note the finer
    vertical scale in (b)]. Distinct open markers identify the
    curves. Convergence to unity (dashed line) confirms the leading
    and subleading scaling at low temperature, including the
    Sommerfeld coefficient $a_2 = 4\sqrt{2}/(9\pi)$ in
    \eqref{eq:A_tau_lowtau} and the linear law
    $B(\tau)\simeq -(16\sqrt{2}/3\pi^3)\,\tau$ of
    \eqref{eq:Bsad_lowtau_final_closed}. The deviation visible at
    the smallest $N$ and the smallest $\tau$ decreases with $N$ at
    the rate set by the next term of the expansion, and marks the
    approach to the Airy window $\tau\sim N^{-2/3}$ of
    Eq.~\eqref{eq:validity_window}.}
  \label{fig:low_tau}
\end{figure}

% =====================================================================
\subsubsection{High temperature}
\label{sec:Bsad_hightau}
% =====================================================================

For $\tau\gg 1$ the saddle fugacity is small, $z=e^{\beta\mu}\ll 1$,
and the Fermi factor \eqref{eq:f_tau} is uniformly dilute in the
relevant phase-space region. To leading order in the fugacity,
\begin{equation}
f_\tau(1-f_\tau)\simeq f_\tau,
\qquad
f_\tau(1-f_\tau)(1-2f_\tau)\simeq f_\tau,
\label{eq:boltzmann_approximations}
\end{equation}
so $J_\alpha\simeq I_\alpha$ and $K_\alpha\simeq I_\alpha$.
Combined with \eqref{eq:g0_t}--\eqref{eq:g0_tt} this gives
\begin{equation}
\partial_t g_0 \simeq 2 g_0,\qquad
\partial_t^2 g_0 \simeq 4 g_0,
\label{eq:dtg_boltz}
\end{equation}
and at the global level $V_3(\tau)\simeq V_2(\tau)$. Substituting
into \eqref{eq:H_def},
\begin{equation}
\mathcal H(u;\tau)
\simeq \frac{1}{2} \left[\frac{4 g_0}{V_2}-\frac{2 g_0}{V_2}\right]
=\frac{g_0(u;\tau)}{V_2(\tau)},
\label{eq:H_boltz_simplified}
\end{equation}
so that, by \eqref{eq:Bsad_master} and the definition
\eqref{eq:A_tau_def} of $A(\tau)$,
\begin{equation}
B(\tau)\simeq -\pi\,\frac{A(\tau)}{V_2(\tau)}.
\label{eq:Bsad_A_over_V2}
\end{equation}

In the dilute regime the grand-canonical particle-number distribution
becomes Poissonian, so $\kappa_2\simeq\langle N\rangle$, which through
\eqref{eq:kappa2_scaling} fixes
\begin{equation}
V_2(\tau)\xrightarrow[\tau\to\infty]{}\pi.
\label{eq:V2_to_pi}
\end{equation}
Combining \eqref{eq:Bsad_A_over_V2}, \eqref{eq:V2_to_pi} with the
high-$\tau$ leading asymptote of $A(\tau)$ from
\eqref{eq:A_hightau_final},
\begin{equation}
B(\tau)\xrightarrow[\tau\to\infty]{}-A(\tau)
=-\frac{\sqrt{\tau}}{\pi^{3/2}},
\label{eq:Bsad_hightau_leading}
\end{equation}
so $\mathcal C_N(\tau)\sim (\sqrt\tau/\pi^{3/2})\,N^{5/2}
-(\sqrt\tau/\pi^{3/2})\,N^{3/2}+\cdots$. Thus, in the Boltzmann
regime, the canonical $N^{3/2}$ correction is equal in magnitude and
opposite in sign to the leading $N^{5/2}$ coefficient---a universal
ratio $B/A\to-1$ that we revisit in Sec.~\ref{sec:GCE} as a direct
manifestation of the canonical-vs-grand-canonical fluctuation
mismatch.

The first $1/\tau$ correction follows from extending the virial
expansion \eqref{eq:f_virial_A} of $f_\tau$ to next-to-leading order
in the fugacity, in parallel with the derivation of
\eqref{eq:A_hightau_final}. The calculation is given in
Appendix~\ref{app:asymptotics_B_high} and yields
\begin{equation}
B(\tau)=-\frac{\sqrt{\tau}}{\pi^{3/2}}
 \left[1+\frac{5-3\sqrt 3}{2}\,\frac{1}{\tau}
+ \mathcal O \left(\frac{1}{\tau^2}\right)\right],
\qquad (\tau\gg 1),
\label{eq:Bsad_hightau_final_1overTau}
\end{equation}
where the coefficient of the $1/\tau$ term is
negative, so this correction tempers the magnitude of
$B(\tau)$ relative to its leading Boltzmann form
$-\sqrt\tau/\pi^{3/2}$.

\begin{figure}[h!]
  \centering
  \includegraphics[width=\textwidth]{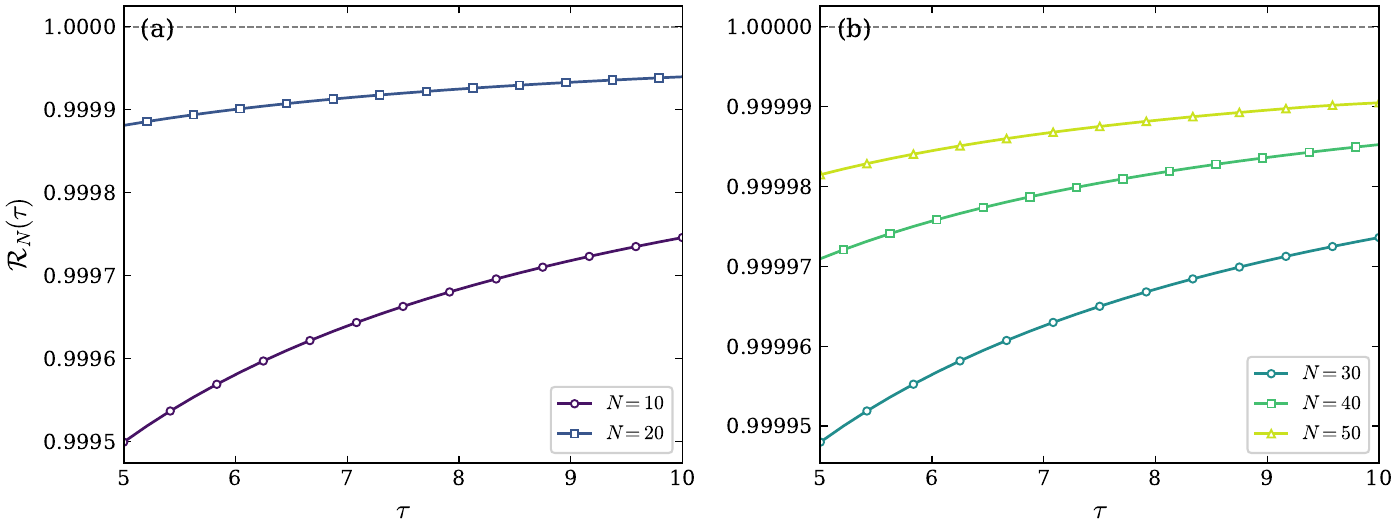}
  \caption{High-$\tau$ verification of the canonical scaling law: the
    same ratio $\mathcal R_N$ as in Fig.~\ref{fig:low_tau}, on
    $\tau\in[5,10]$: (a) $N=10,20$; (b) $N=30,40,50$. Distinct open
    markers identify the curves. Every plotted point satisfies the
    spatial-truncation criterion \eqref{eq:MMAX_safety} with margin
    $\geq1.2$ and uses the underflow-safe evaluation of
    Appendix~\ref{app:numerical}. The
    collapse onto unity (dashed line) improves monotonically with
    $N$ and confirms the asymptotic forms $A(\tau)$ from
    \eqref{eq:A_hightau_final} and $B(\tau)$ from
    \eqref{eq:Bsad_hightau_final_1overTau} in the Boltzmann regime,
    including the universal ratio $B/A\to-1$ of
    \eqref{eq:Bsad_hightau_leading}.}
  \label{fig:high_tau}
\end{figure}

\subsection{Numerical evaluation for intermediate $\tau$}
\label{sec:numerical_AB}
In the intermediate regime
$\tau\sim 1$ no small parameter is available, so the scaling functions
$A(\tau)$ and $B(\tau)$ must be computed numerically from their integral
representations \eqref{eq:A_tau_def} and \eqref{eq:Bsad_oneline}.

The procedure is straightforward in principle: for each $u$, we evaluate
the six $q$-integrals \eqref{eq:I0_I2_def}, \eqref{eq:J0_J2},
\eqref{eq:K0_K2}; form $g_0(u;\tau)$, $\partial_t g_0(u;\tau)$ and
$\partial_t^2 g_0(u;\tau)$ via \eqref{eq:g0_t}--\eqref{eq:g0_tt}; and
integrate in $u$ to obtain $A(\tau)$, $V_2(\tau)$, $V_3(\tau)$ and
$B(\tau)$ through \eqref{eq:Bsad_oneline}. The Fermi factor is
evaluated with the self-consistent $\xi(\tau)$ from \eqref{eq:xi_eq};
all integrands decay exponentially at large argument, so
truncating each integration domain at $\sim \sqrt{\tau}$ times an
$\mathcal O(1)$ factor introduces only exponentially small deviation. The full
quadrature scheme, convergence criteria, and a discussion of error
control for the partial cancellations in $\mathcal H(u;\tau)$ are
documented in Appendix~\ref{app:numerical}.

\begin{figure}[h!]
  \centering
  \begin{subfigure}[t]{0.49\textwidth}
    \centering
    \includegraphics[width=\textwidth]{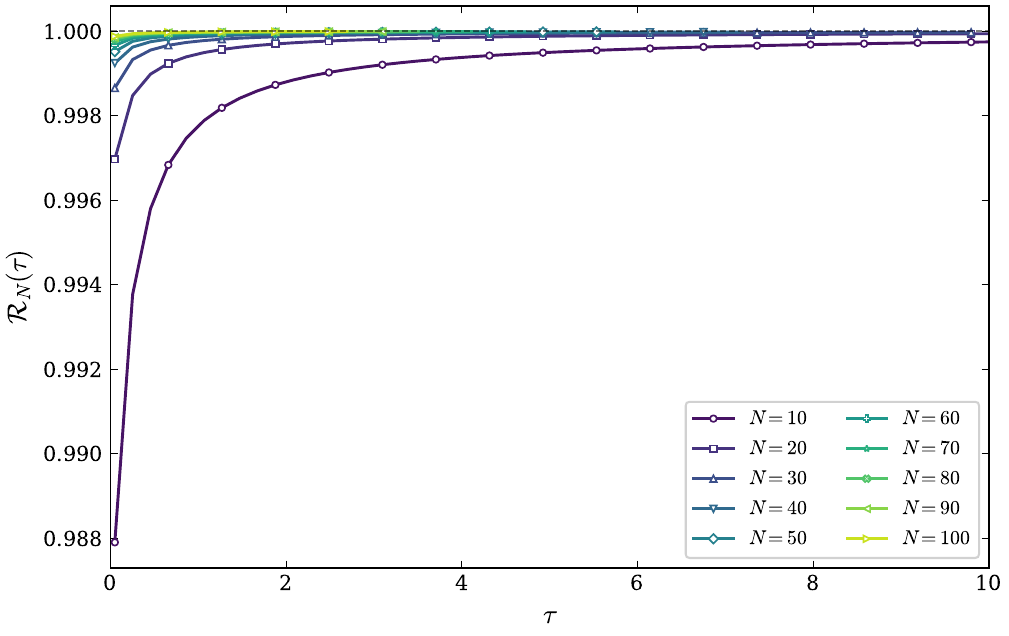}
    \caption{}
    \label{fig:full_tau}
  \end{subfigure}
  \hfill
  \begin{subfigure}[t]{0.49\textwidth}
    \centering
    \includegraphics[width=\textwidth]{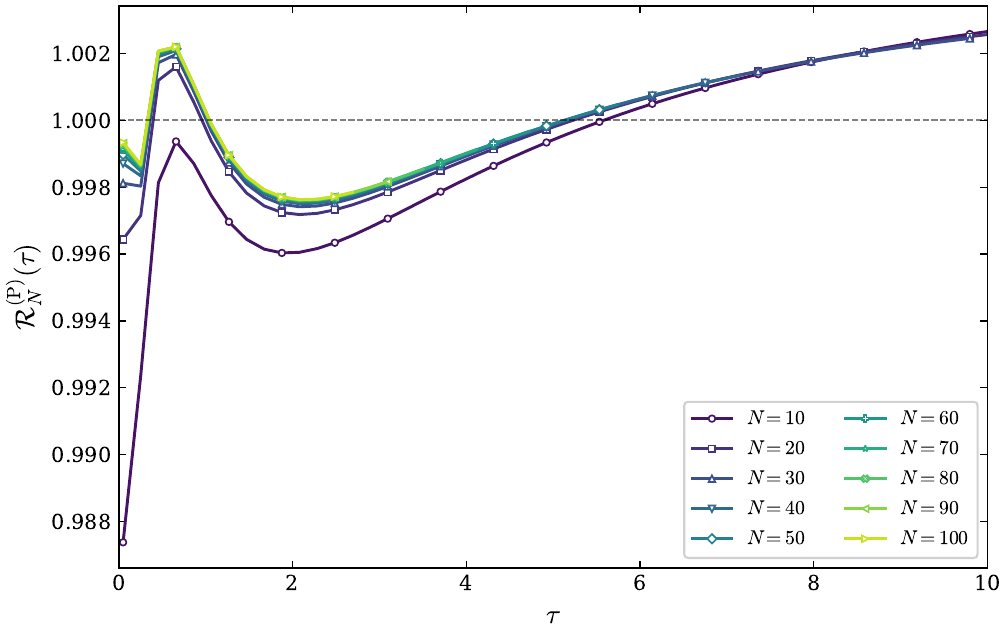}
    \caption{}
    \label{fig:pade_ratio}
  \end{subfigure}
  \caption{Full-$\tau$ scaling verification of Eq.~\eqref{eq:scaling_law}
    across $\tau\in[0,10]$ and $N=10,\ldots,100$. (a) Ratio
    $\mathcal R_N$ from \eqref{eq:RN_def}, evaluated with the numerical
    scaling functions $A(\tau)$ and $B(\tau)$. (b) Ratio
    $\mathcal R_N^{(\mathrm{P})}$ from \eqref{eq:RNP_def}, evaluated
    with the Pad\'e approximants $A_{\mathrm{P}}(\tau)$ and
    $B_{\mathrm{P}}(\tau)$ of \eqref{eq:A_pade} and \eqref{eq:B_pade}.
    In both panels the curves collapse onto unity (dashed line),
    confirming the canonical scaling law from the degenerate
    ($\tau\ll 1$) to the Boltzmann ($\tau\gg 1$) regimes and the
    validity of the Pad\'e approximants as closed-form substitutes for
    the universal integrals. In (a) the residual shrinks with $N$,
    as the $\mathcal O(N^{1/2})$ next-to-subleading correction
    requires, whereas in (b) it saturates at an $N$-independent
    level set by the accuracy of the approximants themselves,
    Eqs.~\eqref{eq:A_pade_error} and \eqref{eq:B_pade_error}. Each
    curve is truncated at the largest $\tau$ satisfying the
    spatial-truncation criterion \eqref{eq:MMAX_safety} for
    $M_{\max}=5000$, which is why the higher-$N$ curves span a
    shorter range.}
  \label{fig:scaling_verification}
\end{figure}

\begin{figure}[h!]
  \centering
  \includegraphics[width=\textwidth]{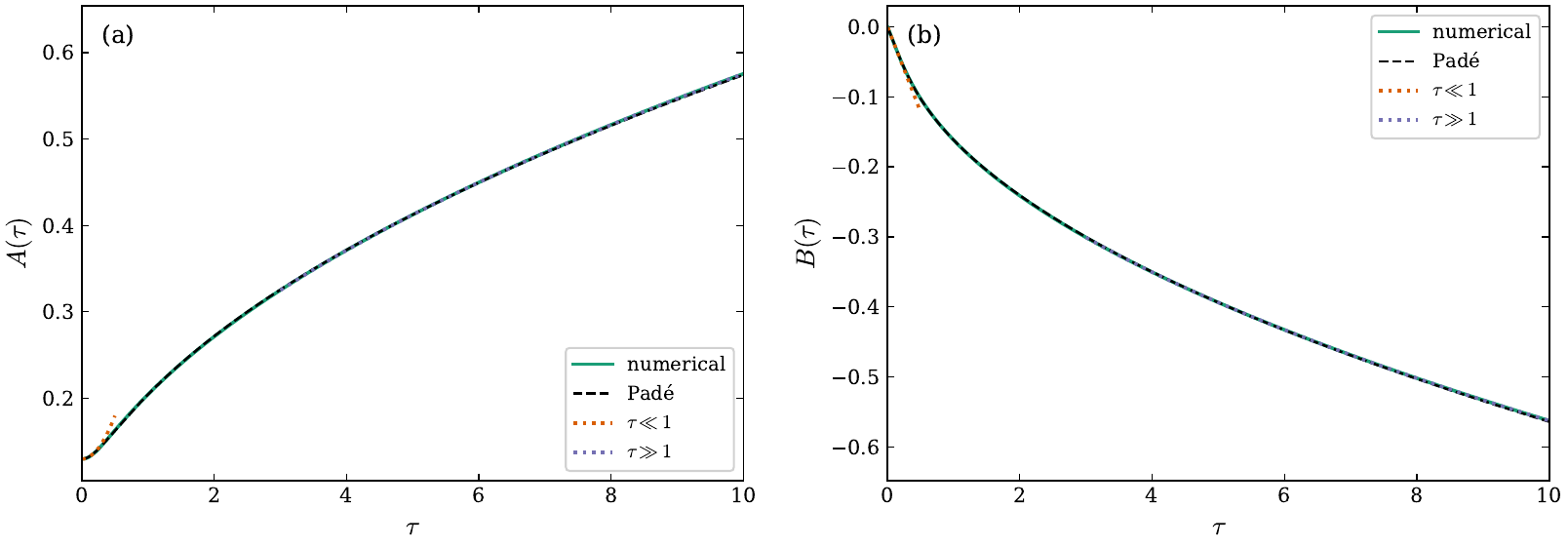}
  \caption{Universal scaling functions $A(\tau)$ (left panel) and
    $B(\tau)$ (right panel). Solid blue: numerical evaluation of the
    integral representations \eqref{eq:A_tau_def} and
    \eqref{eq:Bsad_oneline} using the self-consistent $\xi(\tau)$ from
    \eqref{eq:xi_eq}. Dashed black: Pad\'e approximants
    \eqref{eq:A_pade} and \eqref{eq:B_pade}, indistinguishable
    from the numerical curves on this scale. Dotted red: low-$\tau$ Sommerfeld expansions
    \eqref{eq:A_tau_lowtau} and \eqref{eq:Bsad_lowtau_final_closed},
    shown on $\tau\le0.5$; the latter is the purely linear law, the
    $\tau^{2}$ edge contributions having cancelled.
    Dotted green: high-$\tau$ virial expansions
    \eqref{eq:A_hightau_final} and
    \eqref{eq:Bsad_hightau_final_1overTau}, shown on
    $\tau\ge3$.}
  \label{fig:AB_universal}
\end{figure}

To validate the universal-integral evaluation, we compute the canonical
contact $\mathcal C_N(\tau)$  by contour projection and
check that \eqref{eq:scaling_law} holds with coefficients consistent
with \eqref{eq:A_tau_def} and \eqref{eq:Bsad_oneline}. Concretely, for
fixed $\tau$ and a set of particle numbers $N\in\{N_1,\dots,N_M\}$ we fit
\begin{equation}
\frac{\mathcal C_N(\tau)}{N^{5/2}} = A(\tau)+\frac{B(\tau)}{N}+\mathcal O\left(N^{-2}\right),
\label{eq:CE_fit_formula}
\end{equation}
so that a linear regression of $\mathcal C_N(\tau)/N^{5/2}$ against
$1/N$ returns $A(\tau)$ as the intercept and $B(\tau)$ as the slope.
Agreement with the universal-integral values serves as a strong
consistency check for both the scaling exponents and the numerical
implementation. Compact closed-form Pad\'e approximants
$A_{\mathrm P}(\tau)$ and $B_{\mathrm P}(\tau)$, valid uniformly on
the reduced temperature range, are constructed in Appendix~\ref{app:numerical} ---
eqs.~\eqref{eq:A_pade} and \eqref{eq:B_pade} --- from analytic asymptotic
constraints supplemented by a minimax fit. The scaling-law verification
using both the numerical and the Pad\'e scaling functions is shown in
Fig.~\ref{fig:scaling_verification}, and the resulting $A(\tau)$ and
$B(\tau)$ are displayed in Fig.~\ref{fig:AB_universal} together with
their analytic Sommerfeld and virial limits.

We observe that, at low
temperature (Fig.~\ref{fig:low_tau}),  the ratio $\mathcal R_N$
approaches unity as $N$ grows, which tests the two scaling
exponents and the Sommerfeld forms \eqref{eq:A_tau_lowtau} and
\eqref{eq:Bsad_lowtau_final_closed} together. The residual grows
towards the left edge of the window and has the largest  value for the smallest
particle number. Moreover, it shrinks with $N$ at the rate expected of
the next term in the series. Such a deviation is a signal of its boundary: at fixed $N$, lowering
$\tau$ eventually reaches the Airy window
$\tau\sim N^{-2/3}$ of Eq.~\eqref{eq:validity_window}, where the
thermal edge is no longer the widest scale and the fixed-$\tau$
ordering breaks down. At high temperature
(Fig.~\ref{fig:high_tau}) the collapse is cleaner, and it confirms
the virial forms \eqref{eq:A_hightau_final} and
\eqref{eq:Bsad_hightau_final_1overTau} together with the Boltzmann
ratio $B/A\to-1$ of \eqref{eq:Bsad_hightau_leading}. Every point
plotted there satisfies the truncation criterion
\eqref{eq:MMAX_safety}.

Across the whole temperature range
(Figs.~\ref{fig:scaling_verification} and \ref{fig:AB_universal})
the same collapse holds with the numerical scaling functions, the
deviation again falling with $N$ like the next-to-subleading term.
Replacing the universal integrals by the Pad\'e approximants leaves
the collapse intact but changes the character of the residual: it
stops shrinking with $N$ and settles at the level set by the
accuracy of the approximants themselves,
Eqs.~\eqref{eq:A_pade_error} and \eqref{eq:B_pade_error}. Its
independence of $N$ is what identifies its origin. For applications
in which closed forms are more convenient than quadrature, the
approximants may therefore stand in for
\eqref{eq:A_tau_def} and \eqref{eq:Bsad_oneline} throughout the
range considered.

%====================%====================%====================
\section{Ensemble correspondence}
\label{sec:GCE}
%====================%====================%====================

The canonical analysis of Sec.~\ref{sec:CE_leading_scaling}--\ref{sec:saddle_Btau}
established the scaling law \eqref{eq:scaling_law}
through a contour-integral representation followed by saddle reduction.
The same machinery, evaluated at the saddle alone, is essentially the
grand-canonical calculation. Examining what survives---and what
disappears---when the contour is removed clarifies the physical origin
of the subleading coefficient $B(\tau)$ as an ensemble-correspondence
effect.

In the GCE the occupations $\eta_n\in\{0,1\}$ are independent Bernoulli
variables with means
$\bar n_n(z)=\left( e^{\beta(\varepsilon_n-\mu)}+1 \right)^{-1}$ and joint moments
$\langle\eta_m\eta_n\rangle_{\GCE}=\bar n_m\bar n_n$ for $m\ne n$.
Inserting this factorisation into the pair representation of the
integrand (cf.\ Appendix~\ref{app:kernel}) and using $A_{nn}(x):= 0$ collapses the
double sum to the same kernel functional that appears under the
contour in \eqref{eq:FN_kernel_contour_final}, but evaluated directly
at the GCE fugacity $z=e^{\beta\mu}$:
\begin{equation}
F^{\GCE}(x;z)=\rho_z(x)\,\kappa_z(x)-S_z(x)^2,
\qquad
\mathcal C^{\GCE}=\frac{2}{\pi}\int dx\,F^{\GCE}(x;z),
\label{eq:FGCE_kernel}
\end{equation}
with $\rho_z,\kappa_z,S_z$ defined as in \eqref{eq:local_defs}. No
contour integration is required: the right-hand side of
\eqref{eq:FGCE_kernel}  {is} the GCE expectation.

To match $\langle N\rangle_{\GCE}=N$ we impose the same number condition
\eqref{eq:saddle_number} as the canonical saddle, which in the
fixed-$\tau$ scaling limit (with $\beta=1/(\tau N)$ and $\mu=\xi(\tau)\,N$)
reduces to the self-consistent transcendental equation \eqref{eq:xi_eq}
already introduced in Sec.~\ref{sec:CE_leading_scaling}. The
phase-space Fermi factor is the one in \eqref{eq:f_tau}, evaluated with
the self-consistent $\xi(\tau)$.

The LDA reduction of \eqref{eq:FGCE_kernel} is identical to that
performed in Sec.~\ref{sec:CE_leading_scaling}; setting $S_z:= 0$ by
parity and using the scaling variables \eqref{eq:scaling_xp},
\begin{equation}
F^{\GCE}(x;z)=\frac{N^2}{\pi^2}\,g_0(u;\tau)+\mathcal O(1),
\qquad x=\sqrt{2N}\,u,
\label{eq:FGCE_no_N32}
\end{equation}
with $g_0(u;\tau)=I_0(u;\tau)\,I_2(u;\tau)$ as in
\eqref{eq:Gstar_scaling_g0}. The crucial difference with respect to
the canonical decomposition \eqref{eq:FN_saddle} is the absence
of the $\mathcal O(N)$ saddle-fluctuation term: \eqref{eq:FGCE_kernel} contains
no contour integral, and therefore no Gaussian fluctuations to expand
around. Integrating \eqref{eq:FGCE_no_N32} yields
\begin{equation}
\;\mathcal C^{\GCE}_{\langle N\rangle = N}(\tau)
=A(\tau)\,N^{5/2}+\mathcal O(N^{1/2}),\;
\label{eq:CGCE_full}
\end{equation}
with the  {same} $A(\tau)$ as in \eqref{eq:A_tau_def}, now
understood with the self-consistent $\xi(\tau)$, and  {no}
$N^{3/2}$ correction at the LDA order. The leading scaling
function $A(\tau)$ is therefore an ensemble-independent universal
quantity.

It is important to distinguish two senses in which
\eqref{eq:CGCE_full} holds. At the LDA level, the absence of
$N^{3/2}$ corrections is true  {by construction}: the only $N$-dependence
in $F^{\GCE}(x;z_\star)$ enters through $\sqrt{2N}$ in the
rescaled coordinate and through the Fermi factor at the
LDA-saddle fugacity, both of which produce only even powers of
$N^{1/2}$. Beyond the LDA level, the finite-$N$ mechanisms that could
in principle generate an $N^{3/2}$ term in \eqref{eq:CGCE_full} are:
(i) the discreteness of the number equation used to fix the fugacity
at $\langle N\rangle_{\GCE}=N$, whose continuum replacement could
shift the chemical potential; (ii) the Wigner--Kirkwood corrections to the phase-space evaluation of the
kernel functional; and (iii) the kernel term $S_z^{2}$, which
the leading phase-space parity argument sets to zero. The power
counting of Appendix~\ref{app:powercounting} shows that all three
enter only at order $N^{1/2}$: the midpoint structure $\varepsilon_n=n+\frac12$ of the oscillator
spectrum removes the $\mathcal O(1)$ Euler--Maclaurin boundary term
from the number equation, so the induced shift of the scaled
chemical potential is only $\mathcal O(N^{-2})$. The
Wigner--Kirkwood corrections are of relative order $N^{-2}$ at
fixed $\tau$, and the identity $S_z=\frac12\rho_z'$ makes
$S_z^{2}$ a relative-$N^{-2}$ effect as well. Equation \eqref{eq:CGCE_full} therefore holds beyond mere
LDA power counting: the grand-canonical contact at fixed mean
particle number contains no $N^{3/2}$ term, and the interpretation
of $B(\tau)N^{3/2}$ in \eqref{eq:DeltaC} as the complete ensemble
correction is justified.

Subtracting \eqref{eq:CGCE_full} from the canonical expansion
\eqref{eq:scaling_law} yields the central observation of this
section,
\begin{equation}
\Delta \mathcal C(\tau)
:= \mathcal C_N^{\CE}(\tau)-\mathcal C^{\GCE}_{\langle N\rangle = N}(\tau)
=B(\tau)\,N^{3/2}+\mathcal O(N^{1/2}),\;
\label{eq:DeltaC}
\end{equation}
with $B(\tau)$ given explicitly by \eqref{eq:Bsad_oneline}. Equation
\eqref{eq:DeltaC} provides the physical interpretation of the
subleading scaling function: $B(\tau)\,N^{3/2}$ is the
 {ensemble correction} to the contact---the part of $\mathcal C_N^{\CE}$
sourced by the constraint of fixed $N$ and therefore absent from the
GCE. Equivalently, the GCE delivers the leading $A(\tau)\,N^{5/2}$
directly from the local kernel functional, while the CE pays the
price of a fluctuation correction of magnitude exactly
$B(\tau)\,N^{3/2}$. 

The result \eqref{eq:DeltaC} can be derived independently from the
standard ensemble-correspondence formula (see e.g.\ the textbook
discussion in \cite{Kubo1965} or the modern review of
large-deviation methods in statistical mechanics 
\cite{Touchette2009}): for any extensive observable $O$,
\begin{equation}
\langle  O\rangle_{\CE}-\langle  O\rangle_{\GCE}
\simeq
\frac{1}{2\kappa_2}\,\partial_N^2\langle  O\rangle_{\GCE}
-\frac{\kappa_3}{2\kappa_2^{\,2}}\,\partial_N\langle O\rangle_{\GCE},
\label{eq:ensemble_correction}
\end{equation}
with $\kappa_2,\kappa_3$ the GCE particle-number cumulants. In the
fixed-$\tau$ scaling limit, the cumulants reduce as in
\eqref{eq:kappa2_scaling}--\eqref{eq:kappa3_scaling}, exhibiting the
global functions $V_2,V_3$ as nothing more than rescaled
particle-number cumulants of the GCE. Setting $  O= \mathcal C$ with
$\langle \mathcal C\rangle_{\GCE}=A(\tau)\,N^{5/2}$, each derivative drops a
power of $N$ while each $1/\kappa$ factor restores $1/N$, so both
terms in \eqref{eq:ensemble_correction} contribute at order $N^{3/2}$.
Reconstructing the coefficients with the explicit Wigner integrands
recovers \eqref{eq:Bsad_oneline} term by term, and confirms that the
structure of $B(\tau)$---a local part proportional to
$\partial_t^2 g_0/V_2$ and a global part proportional to
$V_3/V_2^{\,2}$---is dictated by the curvature and skewness of the GCE
particle-number distribution.

In the dilute Boltzmann regime $\tau\gg 1$, the GCE distribution of
$N$ becomes Poissonian, $\kappa_2\simeq\langle N\rangle$ and hence
$V_2(\tau)\to\pi$ as in \eqref{eq:V2_to_pi}. The ratio $B/A$ then
collapses to the universal value $-1$ as in
\eqref{eq:Bsad_hightau_leading}, in agreement with
\eqref{eq:Bsad_hightau_final_1overTau}.

%====================%====================%====================
\section{Experimental implications}
\label{sec:experimental}
%====================%====================%====================

The scaling law \eqref{eq:scaling_law} is directly relevant to
trapped-gas experiments, for two reasons. Firstly, an isolated tube
holds a fixed number of atoms for the duration of a shot: the
canonical ensemble is the
statistically appropriate description of the samples on which the
contact is measured. Secondly, at the mesoscopic particle numbers of
current one-dimensional experiments the two terms of
\eqref{eq:scaling_law} are separated by only one power of $N$, so
the ensemble correction is not an academic refinement. In this
section we quantify where it becomes visible.

\subsection{Size of the ensemble correction and where it is visible}
\label{sec:exp_magnitude}
The relative weight of the subleading term is
\begin{equation}
r_N(\tau):=\frac{\left|B(\tau)\right| N^{3/2}}{A(\tau)\,N^{5/2}}
=\frac{1}{N}\left|\frac{B(\tau)}{A(\tau)}\right| ,
\label{eq:rN_def}
\end{equation}
where the universal ratio $|B/A|$ grows monotonically with
temperature, from $\frac{15}{8}\tau$ in the degenerate regime
[Eqs.~\eqref{eq:A_tau_lowtau} and
\eqref{eq:Bsad_lowtau_final_closed}] to $1$ in the Boltzmann limit
[Eq.~\eqref{eq:Bsad_hightau_leading}]. The correction is therefore
largest for small, hot clouds and is bounded there by $1/N$, the
saturation reflecting the Poissonian number statistics of the
dilute gas. Table~\ref{tab:rN} gives $r_N(\tau)$ over the
experimentally relevant window, where it ranges from a fraction of
a percent for the largest clouds considered to the several-percent
level for a tube of ten atoms. 

\begin{table}[t]
\centering
\caption{Relative magnitude $r_N(\tau)=|B(\tau)|N^{3/2}/[A(\tau)N^{5/2}]$
of the canonical-ensemble correction to the contact, in percent,
from the numerical evaluation of the universal scaling functions
\eqref{eq:A_tau_def} and \eqref{eq:Bsad_oneline}.}
\label{tab:rN}
\begin{tabular}{lcccc}
\toprule
 & $N=10$ & $N=25$ & $N=50$ & $N=100$ \\
\midrule
$\tau=0.1$  & $1.8\%$ & $0.74\%$ & $0.37\%$ & $0.18\%$ \\
$\tau=0.25$ & $4.2\%$ & $1.7\%$  & $0.83\%$ & $0.42\%$ \\
$\tau=0.5$  & $6.3\%$ & $2.5\%$  & $1.3\%$  & $0.63\%$ \\
$\tau=1$    & $7.9\%$ & $3.2\%$  & $1.6\%$  & $0.79\%$ \\
$\tau=2$    & $8.9\%$ & $3.6\%$  & $1.8\%$  & $0.89\%$ \\
$\tau=5$    & $9.6\%$ & $3.8\%$  & $1.9\%$  & $0.95\%$ \\
\bottomrule
\end{tabular}
\end{table}

Since $B(\tau)<0$ for all $\tau>0$, the canonical contact always
lies below the grand-canonical (LDA) prediction evaluated at
the same mean particle number: an analysis of fixed-$N$ samples
based on the leading term alone systematically overestimates
the contact by $r_N(\tau)$.

Resolving the correction requires a relative accuracy on
$\mathcal C$ better than $r_N$. Since $r_N=|B/A|/N$ and $|B/A|$ is
bounded by unity, a measurement with fractional resolution
$\epsilon$ can reach the correction only for
$N\lesssim|B/A|/\epsilon$, a bound that tightens as the gas is
made colder and $|B/A|$ falls. Percent-level resolution therefore
brings tubes of a few tens of atoms into range, and the
requirement is least demanding in the nondegenerate regime, where
$|B/A|$ saturates.

The recent direct measurement of the contact in a Lieb--Liniger
gas \cite{Huang2025} is instructive on this point. It was
performed on an array of tubes with a few tens of atoms each, at
intermediate coupling, and reached an accuracy at the ten-percent
level. For those tube occupancies Table~\ref{tab:rN} places the
ensemble correction an order of magnitude below the quoted
resolution. The smaller tubes of lattice-based Tonks--Girardeau
experiments \cite{Paredes2004,Kinoshita2004} are the more
favourable case, since $r_N$ grows as the occupancy falls.

In a lattice-based measurement the observed signal is summed over
an array of tubes with a distribution of atom numbers $\{N_t\}$.
Because each tube is number conserving, the canonical form is the
correct per-tube building block,
\begin{equation}
\mathcal C_{\mathrm{tot}}
=\sum_{t}\left[A(\tau_t)\,N_t^{5/2}+B(\tau_t)\,N_t^{3/2}\right],
\qquad \tau_t=\frac{T}{N_t\hbar\omega},
\label{eq:tube_sum}
\end{equation}
for an array in global thermal equilibrium at temperature $T$.
Since the correction enters with weight $N_t^{3/2}$, the array
average of $r_N$ is dominated by the smaller tubes of the
distribution and generically exceeds $r_{\bar N}$ evaluated at the
mean atom number.

\subsection{Contact thermometry at fixed particle number}
\label{sec:exp_thermo}
The contact is an established thermometer for strongly interacting
1D gases \cite{Yao2018,Huang2025}: one measures $\mathcal C$ and
inverts the theoretical curve $\mathcal C(\tau)$ at known $N$. The
scaling law \eqref{eq:scaling_law} quantifies the systematic error
incurred when the inversion uses the grand-canonical (LDA) curve
$A(\tau)N^{5/2}$ alone. In the Sommerfeld window, where
$A(\tau)=A(0)\left[1+\frac{5\pi^{2}}{32}\tau^{2}\right]$ and
$B(\tau)\simeq b_1\tau$ with $b_1=-16\sqrt2/(3\pi^{3})$, the two
temperature dependences combine into a $\tau$-independent shift of
the inferred reduced temperature,
\begin{equation}
\delta\tau
:=\tau_{\mathrm{fit}}-\tau
\simeq-\frac{|b_1|}{2a_2\,N}
=-\frac{6}{\pi^{2}N},
\label{eq:thermometry_bias}
\end{equation}
with $a_2=4\sqrt2/(9\pi)$ the Sommerfeld coefficient of
\eqref{eq:A_tau_lowtau}: a grand-canonical contact thermometer
reads systematically cold by $|\delta\tau|\simeq0.61/N$,
independently of the temperature being measured. At a few tens of
atoms the shift is a sizeable fraction of the reduced temperatures
typical of degenerate samples, and it grows once the Sommerfeld
window is left. Against the temperature accuracies demonstrated by
1D thermometry \cite{Jacqmin2011,Vogler2013}, the canonical
correction is a leading systematic at mesoscopic $N$.

This also quantifies how the thermal broadening of the Fermi
surface interplays with finite $N$ in the interpretation of data.
The thermal correction to the contact itself is
$\mathcal O(\tau^{2})$, while the ensemble correction is linear in
$\tau$ and $1/N$-suppressed. The two are comparable along
$\tau\sim1/N$, so both must be retained for quantitative work
below $\tau\sim1$. The analysis applies within the window
\eqref{eq:validity_window}, $\tau\gg N^{-2/3}$, where thermal
smearing dominates the quantum (Airy) width of the Fermi edge, a
condition that becomes restrictive at small $N$. Colder samples at mesoscopic $N$ leave the
thermal regime altogether and should instead be analysed with the
zero-temperature finite-$N$ expansions of
Refs.~\cite{VignoloMinguzzi2013,Rizzi2018}, whose leading edge
corrections scale as $N^{3/4}$.

\subsection{Relation to existing data and outlook}
\label{sec:exp_outlook}
A direct confrontation of \eqref{eq:scaling_law} with the data of
Ref.~\cite{Huang2025} is not yet possible. That experiment
operates at intermediate coupling, whereas the present analysis
holds in the Tonks--Girardeau limit, and its resolution is not yet
at the level the correction requires. What the scaling law does
provide are two predictions that a Tonks--Girardeau sample with
calibrated atom number can test. The per-tube contact should fall
below the local-density prediction, by the amount tabulated and
with the temperature dependence of $r_N(\tau)$ shown there, and
contact thermometry should carry the $N$-dependent offset
\eqref{eq:thermometry_bias}. Extending $B(\tau)$ to finite
coupling, along the lines of the finite-$\gamma$ analysis of
Ref.~\cite{Yao2018}, is what would make a comparison at
$\gamma\simeq1$ possible.

\section{Conclusions}
\label{sec:conclusions}

We have derived the canonical-ensemble large-$N$ scaling law
\eqref{eq:scaling_law} of Tan's contact for the harmonically
trapped Tonks--Girardeau gas at fixed reduced temperature.
The leading coefficient $A(\tau)$ coincides with the
local-density-approximation result and reproduces the known
zero-temperature value $A(0)=128\sqrt2/(45\pi^3)$. The subleading
coefficient $B(\tau)$, the central new object of this work, admits a
universal first-principles representation \eqref{eq:Bsad_oneline} in
terms of phase-space integrals of the Fermi factor, has explicit
Sommerfeld and virial limits given in
\eqref{eq:Bsad_lowtau_final_closed} and
\eqref{eq:Bsad_hightau_final_1overTau}   and is identified through
\eqref{eq:DeltaC} with the canonical-versus-grand-canonical ensemble
difference at fixed mean particle number. In the Boltzmann limit, the
universal asymptotic coefficient ratio $B/A\to -1$ — equivalently
$B(\tau\gg 1)\simeq -A(\tau)$ — emerges from the Poissonian
particle-number statistics of the GCE: the canonical contact lies
 {below} the corresponding grand-canonical contact at the
canonical-saddle fugacity, with the relative correction
$|B/A|/N = 1/N$ in the dilute classical regime. We construct compact closed-form
Pad\'e approximants for $A(\tau)$ and $B(\tau)$, valid uniformly on
$\tau\in[0,10]$, and the
scaling law is verified numerically against canonical
contour-integration data for $N$ up to $100$ across the full
temperature range.

The expansion derived in this manuscript is for fixed positive $\tau$ followed by $N\to\infty$. As discussed in
Sec.~\ref{sec:Bsad_lowtau}, it holds in the window
$\tau\gg N^{-2/3}$ of Eq.~\eqref{eq:validity_window}, where the
thermal smearing of the Fermi edge dominates its quantum (Airy)
width, and it does not commute with the strict $\tau\to0$ limit,
whose distinct $N^{3/4}$ and $N^{1/4}$ edge hierarchy
\cite{VignoloMinguzzi2013,Rizzi2018} takes over below the
crossover.

The result extends the canonical analysis of
Ref.~\cite{SantAna2019} to many particles through the kernel representation 
and  identifies the precise origin of the subleading $N^{3/2}$ term as a finite-$N$
ensemble-correspondence effect. In light of recent direct contact
measurements in trapped one-dimensional Bose gases
\cite{Huang2025}, the explicit scaling functions $A(\tau)$ and
$B(\tau)$ provide a quantitative target against which
finite-temperature trapped-geometry data can be compared. The
corresponding experimental windows, the expected percent-level
magnitudes, and the systematic bias of grand-canonical contact
thermometry are quantified in Sec.~\ref{sec:experimental}.

Several extensions are natural. The boundary-layer (Airy) corrections
to the contact, which produce additional subleading terms scaling as
$N^{3/4}$ and $N^{1/4}$ at zero temperature
\cite{Rizzi2018}, persist at finite temperature but are
expected to be exponentially suppressed by the thermal smoothing of
the Fermi surface; their explicit finite-$T$ form is the natural
sequel to the present work. The same contour-integral framework can
also be applied to multi-component fermionic mixtures
\cite{Decamp2016,Decamp2018} and to the trapped Lieb--Liniger gas at
finite coupling \cite{Yao2018}, where similar
canonical-vs-grand-canonical distinctions are expected to control the
subleading scaling.

\section*{Data availability}
The numerical data and code that support the findings of this study 
are available at \\ 
\href{https://github.com/ftahas/Contact_Scaling}{github.com/ftahas/Contact\_Scaling}.

%%%%%%%%%%%%%%%%%%APPENDIX%%%%%%%%%%%%%%%%%%%%%%%%%%%%%%%%%%%%%%%%%%%%%%%%%%%%%
\appendix

%%%%%%%%%%%%%%%%%%%%%%%%%%CONTACT%%%%%%%%%%%%%%%%%%%%%%%%%%%%%%%%%%%%%%%%%%%%%%%%
\section{Derivation of the contact}
\label{app:derivation}
We start with the  $j$-body density matrix of $N$ particles at temperature $T$, 
\begin{equation}
	\begin{aligned}\label{rhoj_oi}
		\varrho^{(j)}(x_1,\dots,x_j;x'_1,\dots,x'_j) &= \frac{N!}{(N-j)!} Z^{-1}
		\sum_{\alpha} e^{-\beta E_{\alpha}} \int dx_{j+1} \dots dx_N 
		\Psi_{\alpha}^{(b)*}(x_1,\dots,x_N) \\
		&\times\Psi_{\alpha}^{(b)} (x'_1,\dots,x'_{j},x_{j+1},\dots,x_N),
	\end{aligned}
\end{equation}
where $Z = \sum_{\alpha} e^{-\beta E_{\alpha}}$ is the partition function and 
the system's total energy is simply the summation of all the individual single-particle energies, \textit{i.e.},
$E_{\alpha} = \sum_{i=1}^N \varepsilon_{n_i}$, with $\varepsilon_{n_i} = (n_i + 1/2)\hbar \omega$.
Invoking the Bose--Fermi mapping 
\begin{equation}
\label{mapping}
	\Psi_{\alpha}^{(b)}(x_1,\dots,x_N) = \Theta(x_1,\dots,x_N)\Psi_{\alpha}^{(f)}(x_1,\dots,x_N),
\end{equation}
where $\Theta(x_1,\dots,x_N):=\prod_{i<j}\mathrm{sgn}(x_i-x_j)$ is either $+1$ or $-1$, in order to compensate the anti-symmetrization of the fermionic wave function $\Psi^{(f)}_\alpha$, and $\alpha$ is the quantum number describing the particles in a respective set of individual quantum numbers $\{n_1,n_2,\dots,n_N\}$, and the fermionic many-body wave function is given by the Slater determinant 
\begin{equation}\label{slater}
	\Psi_{\alpha}^{(f)}(x_1,\dots,x_N) = \left(N!\right)^{-1/2} \det[\phi_{n_i}(x_j)]_{n_i\in\{n_1,\dots,n_N\};\,x_j\in\{x_1,\dots,x_N\}},
\end{equation}
with $\phi_n(x)$ being the solutions of the harmonic oscillator single-particle solutions, 
eq. \eqref{rhoj_oi} reads
\begin{equation}
        \begin{aligned}\label{rhoj_oi2}
		&\varrho^{(j)}(x_1,\dots,x_j;x'_1,\dots,x'_j) = \frac{N!}{(N-j)!} Z^{-1}
                \sum_{\alpha} e^{-\beta E_{\alpha}} \int dx_{j+1} \dots dx_N
                \Theta(x_1,\dots,x_N) \\ 
		&\times \Psi_{\alpha}^{(f)}(x_1,\dots,x_N)
		\Theta(x'_1,\cdots,x'_j,x_{j+1},\dots,x_N)\Psi_{\alpha}^{(f)}(x'_1,\dots,x'_j,x_{j+1},\dots,x_N).
        \end{aligned}
\end{equation}
Now we turn our focus to the integrand. It is possible to rewrite the product 
of the $\Theta$'s as 
\begin{equation}
	\begin{aligned}
		\Theta(x_1,\dots,x_N)\Theta(x'_1,\dots,x'_j,x_{j+1},\dots,x_N) =&
	\Theta(x_1,\dots,x_j) \Theta(x'_1,\dots,x'_j) \\
		&\times \prod_{i=j+1}^N \prod_{l=1}^{2j} \mathrm{sgn}(x_i-y_l),
	\end{aligned}
\end{equation}
where $y_1\le y_2\le\dots\le y_{2j}$ denotes the increasing
rearrangement of the full set $\{x_1,\dots,x_j,x'_1,\dots,x'_j\}$
(in general the unprimed and primed coordinates interlace, so the
$y_l$ cannot be identified with the $x_i$ and $x'_i$ in any fixed
order).
Now let us consider the set $\mathfrak{S}=(y_1,y_2)\cup(y_3,y_4)\cup\dots\cup(y_{2j-1},y_{2j})$.
It is straightforward to observe that 
\begin{equation} \prod_{i=1}^{2j} \mathrm{sgn}(x-y_i) =
\begin{cases}
	-1 ,      & x \in \mathfrak{S} \\
	+1,     & x \notin \mathfrak{S}
\end{cases}.
\end{equation}
Denoting the number of variables among $x_{j+1},\dots,x_N$ which are in $\mathfrak{S}$ by $M_{\mathfrak{S}}$, we have that
\begin{equation}
                \Theta(x_1,\dots,x_N)\Theta(x'_1,\dots,x'_j,x_{j+1},\dots,x_N)=
	\Theta(x_1,\dots,x_j) \Theta(x'_1,\dots,x'_j) (-1)^{M_\mathfrak{S}}.
\end{equation}
Consequently, eq. \eqref{rhoj_oi2} results in
\begin{equation}
        \begin{aligned}\label{rhoj_oi3}
		&\varrho^{(j)}(x_1,\dots,x_j;x'_1,\dots,x'_j) = \frac{N!}{(N-j)!} Z^{-1}
		\Theta(x_1,\dots,x_j) \Theta(x'_1,\dots,x'_j)
                \sum_{\alpha} e^{-\beta E_{\alpha}} \\
	&\times \int dx_{j+1} \dots dx_N (-1)^{M_\mathfrak{S}}
                \Psi_{\alpha}^{(f)}(x_1,\dots,x_N)
                \Psi_{\alpha}^{(f)}(x'_1,\dots,x'_j,x_{j+1},\dots,x_N).
        \end{aligned}
\end{equation}
Now, considering any integral of the form 
\begin{equation}
	I = \int dx_1 \dots \int (-1)^{M_\mathfrak{S}} f(x_1,\dots,x_j),
\end{equation}
where $M_\mathfrak{S}$ is the number of integration variables inside 
the subdomain $\mathfrak{S}$ and $f$ is a symmetric function, 
it is possible to write 
\begin{equation}
	I = \sum_{m=0}^j \binom{j}{m} (-1)^m 
	\int_\mathfrak{S}dx_1 \dots dx_m
	\int_{\Re-\mathfrak{S}}dx_{m+1} \dots dx_j f(x_1,\dots,x_j).
\end{equation}
Making use of $\int_{\Re-\mathfrak{S}} dx = \int_{\Re} dx - \int_{\mathfrak{S}} dx$,
we have
\begin{equation}\label{acima}
        I = \sum_{m=0}^j \binom{j}{m} (-1)^m
	\sum_{n=0}^{j-m} \binom{j-m}{n} (-1)^n
	\int_\mathfrak{S}dx_1 \dots dx_{m+n}
        \int_{\Re}dx_{m+n+1} \dots dx_j f(x_1,\dots,x_j).
\end{equation}
Performing the summation for $m+n=i$, \eqref{acima} reduces to
\begin{equation}
	I = \sum_{i=0}^j \binom{j}{i} (-2)^i 
	\int_\mathfrak{S} dx_1 \dots dx_i
	\int dx_{i+1}\dots dx_j f(x_1,\dots,x_j).
\end{equation}
Thence, we have that the $j$-body density matrix \eqref{rhoj_oi3}
can be written as
\begin{equation}
        \begin{aligned}\label{rhoj_oi4}
		\varrho^{(j)}(x_1,\dots,x_j;x'_1,\dots,x'_j) =& \frac{N!}{(N-j)!} Z^{-1}
                \Theta(x_1,\dots,x_j) \Theta(x'_1,\dots,x'_j)
                \sum_{\alpha} e^{-\beta E_{\alpha}} \\
		&\times \sum_{i=0}^{N-j} \binom{N-j}{i} (-2)^i \int_\mathfrak{S} 
		dx_{j+1} \dots dx_{j+i} \int dx_{j+i+1}\dots dx_N \\
		&\times\Psi_{\alpha}^{(f)}(x_1,\dots,x_N)
                \Psi_{\alpha}^{(f)}(x'_1,\dots,x'_j,x_{j+1},\dots,x_N).
        \end{aligned}
\end{equation}

The one-body density matrix gives the Lenard series
\begin{equation}
\begin{aligned}
	\varrho^{(1)}(x,x') =& \frac{N}{Z} \sum_{\alpha} e^{-\beta E_{\alpha}} 
	\sum_{j=1}^{N-1}\binom{N-1}{j} (-2)^j [\mathrm{sgn}(x-x')]^j \int_x^{x'} dx_2 \dots dx_{j+1} \\
	&\times \int dx_{j+2}\dots dx_N \Psi_{\alpha}^{(f)}(x,x_2,\dots,x_N) \Psi_{\alpha}^{(f)}(x',x_2,\dots,x_N).
\end{aligned}
\end{equation}
The sum runs from $j=1$: the $j=0$ term would
give the noninteracting fermionic one-body density matrix
$\varrho^{(1)}_f(x,x')$, which is analytic in $x'-x$ at short
distances and therefore contributes neither to the universal
$|x'-x|^3$ behaviour nor to the $k^{-4}$ tail of the momentum
distribution. The contact is sourced entirely by the non-analytic
$|x'-x|^3$ piece of $\varrho^{(1)}$, generated by the
$[\mathrm{sgn}(x-x')]^j$ factors at $j \ge 1$.
Here it is possible to recognize the $j$-body fermionic correlator as
\begin{equation}
	\varrho^{(1)}(x,x') = \sum_{j=1}^{N-1} \frac{(-2)^j}{j!} [\mathrm{sgn}(x-x')]^j \int_x^{x'} dx_2 \dots dx_{j+1} \varrho^{(j+1)}_f(x,x_2,\dots,x_{j+1};x',x_2,\dots,x_{j+1}),
\end{equation}
where
\begin{equation}
\begin{aligned}\label{rhoj}
	\varrho^{(j)}_f(x_1,\dots,x_j;x_1',\dots,x_j') &= \frac{N!}{(N-j)!}Z^{-1} \sum_\alpha e^{-\beta E_{\alpha}} \\
	& \times \int dx_{j+1} \dots dx_N \Psi_\alpha^{(f)}(x_1,\dots,x_N) \Psi_\alpha^{(f)}(x_1',\dots,x_j',x_{j+1},\dots,x_N).
\end{aligned}
\end{equation}

As we are interested in the contact, we are going to restrict ourselves to small distances, $|x'-x| \ll 1$. 
Therefore, we consider only the term $j=1$, because the terms $j>1$ produce 
negligible results in the small distance approximation:
\begin{equation}
	\begin{aligned}
		\varrho^{(1)}(x,x') \underset{x \to x'} \sim & 2 \, \mathrm{sgn}(x'-x) \int_x^{x'} dx_2 \, \varrho^{(2)}_f(x,x_2;x',x_2) \\
		\approx & 2 \, \mathrm{sgn}(x'-x) \varrho^{(2)}_f(x,R;x',R) \delta x,
	\end{aligned}
\end{equation}
where $R:=(x+x')/2$ and $\delta x := x'-x$.

Now we proceed with the explicit evaluation of $\varrho^{(2)}$
making use of \eqref{rhoj} together with \eqref{slater}.

\bigskip
\textit{N=2 particles}
\begin{equation}
\begin{aligned}
	\varrho^{(2)}_f(x,R;x',R)=&Z^{-1} \sum_{n_1,n_2} e^{-\beta(\varepsilon_{n_1}+\varepsilon_{n_2})}
\begin{vmatrix} \phi_{n_1}(x) & \phi_{n_2}(x) \\
\phi_{n_1}(R)& \phi_{n_2}(R)\\
\end{vmatrix}
\begin{vmatrix} \phi_{n_1}(x') & \phi_{n_2}(x') \\
\phi_{n_1}(R)& \phi_{n_2}(R)\\
\end{vmatrix} \\
	=& Z^{-1} \sum_{n_1,n_2} e^{-\beta(\varepsilon_{n_1}+\varepsilon_{n_2})}\left[\phi_{n_1}\left(R-\delta x/2\right)\phi_{n_2}(R)-\phi_{n_2}(R-\delta x/2)\phi_{n_1}(R)\right] \\
&\times \left[\phi_{n_1}(R+\delta x/2)\phi_{n_2}(R)-\phi_{n_2}(R+\delta x/2)\phi_{n_1}(R)\right] \\
	=& Z^{-1} \sum_{n_1,n_2} e^{-\beta(\varepsilon_{n_1}+\varepsilon_{n_2})}\left[ \left(\phi_{n_1} - \frac{\delta x}{2} \partial_R \phi_{n_1} \right)\phi_{n_2}
	- \left(\phi_{n_2} -\frac{\delta x}{2} \partial_R \phi_{n_2} \right)\phi_{n_1}\right]  \\
	&\times \left[\left(\phi_{n_1}+ \frac{\delta x}{2}\partial_R \phi_{n_1} \right) \phi_{n_2} - \left(\phi_{n_2}+ \frac{\delta x}{2}\partial_R \phi_{n_2} \right) \phi_{n_1}\right]  \\
=& Z^{-1} \sum_{n_1,n_2} e^{-\beta(\varepsilon_{n_1}+\varepsilon_{n_2})}\frac{\delta x^2}{4} \\
	&\times \left[(\phi_{n_2}\partial_R \phi_{n_1})^2+(\phi_{n_1}\partial_R \phi_{n_2})^2 - 2 \phi_{n_1} \phi_{n_2} \partial_R \phi_{n_1} \partial_R \phi_{n_2} \right].
\end{aligned}
\end{equation}

\bigskip
\textit{N=3 particles}
\begin{equation}
\begin{aligned}
	\varrho^{(2)}_f(x,R;x',R)=&Z^{-1} \sum_{n_1,n_2,n_3} e^{-\beta(\varepsilon_{n_1}+\varepsilon_{n_2}+\varepsilon_{n_3})} \\
	& \times \int dx_3 \,
\begin{vmatrix} \phi_{n_1}(x) & \phi_{n_2}(x) & \phi_{n_3}(x)\\
\phi_{n_1}(R)& \phi_{n_2}(R)&\phi_{n_3}(R)\\
\phi_{n_1}(x_3)& \phi_{n_2}(x_3)&\phi_{n_3}(x_3)\\
\end{vmatrix} \begin{vmatrix} \phi_{n_1}(x') & \phi_{n_2}(x') & \phi_{n_3}(x')\\
\phi_{n_1}(R)& \phi_{n_2}(R)&\phi_{n_3}(R)\\
\phi_{n_1}(x_3)& \phi_{n_2}(x_3)&\phi_{n_3}(x_3)\\
\end{vmatrix}  \\
	=& Z^{-1} \sum_{n_1,n_2,n_3} e^{-\beta(\varepsilon_{n_1}+\varepsilon_{n_2}+\varepsilon_{n_3})}
\left[\phi_{n_1}^2 \left(\frac{\delta x}{2} \partial_R \phi_{n_2}  \right)^2 +\phi_{n_1}^2 \left(\frac{\delta x}{2} \partial_R \phi_{n_3}  \right)^2 \right. \\
	&\left. +\phi_{n_2}^2 \left( \frac{\delta x}{2} \partial_R \phi_{n_1} \right)^2 +\phi_{n_2}^2 \left(\frac{\delta x}{2} \partial_R \phi_{n_3} \right)^2
+\phi_{n_3}^2 \left(\frac{\delta x}{2}\partial_R \phi_{n_1}  \right)^2 \right.  \\
	&\left. +\phi_{n_3}^2 \left(\frac{\delta x}{2} \partial_R \phi_{n_2} \right)^2  -2 \phi_{n_1} \phi_{n_2} \frac{\delta x^2}{4} \partial_R \phi_{n_1} \partial_R \phi_{n_2}  \right.  \\
	&\left.-2 \phi_{n_1} \phi_{n_3} \frac{\delta x^2}{4} \partial_R \phi_{n_1} \partial_R \phi_{n_3} -2 \phi_{n_2} \phi_{n_3} \frac{\delta x^2}{4} \partial_R \phi_{n_2} \partial_R \phi_{n_3}\right].
\end{aligned}
\end{equation}

In the steps above we have used the differentiation relation
\begin{equation}
	\partial_R \phi(R) = \frac{\phi(R)-\phi(R-\delta x/2)}{\delta x/2},
\end{equation}
and the orthogonality of the $\phi$'s
\begin{equation}
	\int_{-\infty}^{+\infty} dx \, \phi_m(x) \phi_n(x)= \delta_{m,n}.
\end{equation}

Therefore, from the explicit evaluations for $N=2$ and 3 particles, we can generalize the fermionic two-body density matrix for $N$ particles as
\begin{equation}
	\begin{aligned}
		\varrho^{(2)}_f(x,R;x',R) = & \frac{(x'-x)^2}{4} Z^{-1} \sum_{n_1,n_2,\dots,n_N} e^{-\beta \sum_{i=1}^N \varepsilon_{n_i}} \\
	& \times \sum_{j\neq k} \left\{ \left[ \phi_{n_j}(R)\partial_R \phi_{n_k}(R) \right]^2 - \phi_{n_j}(R) \phi_{n_k}(R) \partial_R \phi_{n_j}(R) \partial_R \phi_{n_k}(R)
\right\}.
	\end{aligned}
\end{equation}

Consequently, we have
\begin{equation}
\varrho^{(1)}(x,x') \approx \frac{|x'-x|^3}{3} F(R),
\end{equation}
with the definition
\begin{equation}
\begin{aligned}
	F(R) :=&  Z^{-1} \sum_{n_1,\dots,n_N} e^{-\beta \sum_{i=1}^N \varepsilon_{n_i}}  \sum_{j\neq k} \left\{ \left[ \phi_{n_j}(R)\partial_R \phi_{n_k}(R) \right]^2 
	-\phi_{n_j}(R) \phi_{n_k}(R) \partial_R \phi_{n_j}(R)\partial_R \phi_{n_k}(R)
\right\}.
\end{aligned}
\end{equation}

%%%%%%%%%%%%%MOMENTUM DISTRIBUTION%%%%%%%%%%%%%%%%%%%%%
Now we will use our analysis of the one-body density matrix in order to inspect the momentum distribution
\begin{equation}
n(k)=\frac{1}{2\pi}\int_{-\infty}^{\infty}dx\int_{-\infty}^{\infty}dx'\;
\ee^{ik(x-x')}\rho^{(1)}(x,x').
\end{equation}
Let
\[
R=\frac{x+x'}{2},\qquad s=x-x',\qquad dx\,dx'=dR\,ds,
\]
so that 
\begin{equation}
n(k)=\frac{1}{2\pi}\int_{-\infty}^{\infty} dR\int_{-\infty}^{\infty} ds\;
\ee^{iks}\,\rho^{(1)} \left(R+\frac{s}{2},\,R-\frac{s}{2}\right).
\end{equation}
The short-distance condition, $|s|\to 0$, gives 
\[
\rho^{(1)} \left(R+\frac{s}{2},R-\frac{s}{2}\right)
\simeq \frac{|s|^3}{3}\,F(R).
\]
Hence, for large $|k|$,
\begin{align}
n(k)
&\simeq \frac{1}{2\pi}\int dR\,\frac{F(R)}{3}\int_{-\infty}^{\infty}ds\;\ee^{iks}|s|^3.
\end{align}
Now we make use of the asymptotic behaviour of the Fourier transform of $|x-x_0|^{a-1}f(x)$,
\begin{equation}
\int dx\,\ee^{-ik(x-x_0)}|x-x_0|^{a-1}f(x)
=\frac{2}{k^a}f(x_0)\cos \left(\frac{\pi a}{2}\right)\Gamma(a).
\end{equation}
Taking $a=4$, $f:= 1$, $x_0=0$, we get
\begin{equation}
\int_{-\infty}^{\infty}ds\;\ee^{iks}|s|^3
=\frac{2}{k^4}\cos(2\pi)\Gamma(4)=\frac{12}{k^4}.
\end{equation}
Therefore,
\begin{align}
n(k)
&\simeq \frac{1}{2\pi}\int dR\,\frac{F(R)}{3}\,\frac{12}{k^4}
= \frac{2}{\pi}\frac{1}{k^4}\int_{-\infty}^{\infty}dR\,F(R).
\end{align}
Finally, with $\mathcal C=\lim_{k\to\infty}k^4 n(k)$,
\begin{equation}
\mathcal C=\frac{2}{\pi}\int_{-\infty}^{\infty} dR\,F(R).
\end{equation}

%%%%%%%%%%%%%%%%%%%%%%%%%%%%%%%KERNEL%%%%%%%%%%%%%%%%%%%%%%%%%%%%%%%%%%%%%
\section{Kernel representation}
\label{app:kernel}
Here we derive the kernel representation for the contact. 
For that, we work with our integrand
\begin{align}
F_N(x)
&:=
Z_N^{-1}
\sum_{n_1<\cdots<n_N} e^{-\beta\sum_{i=1}^N \varepsilon_{n_i}}
\sum_{j\neq k}
\left\{
\left[\phi_{n_j}(x)\,\partial_x\phi_{n_k}(x)\right]^2
-
\phi_{n_j}(x)\phi_{n_k}(x)\,\partial_x\phi_{n_j}(x)\partial_x\phi_{n_k}(x)
\right\},
\label{eq:FN_original}
\end{align}
with the canonical partition function 
\begin{equation}
Z_N = \sum_{n_1<\cdots<n_N} \exp\left(-\beta\sum_{i=1}^N \varepsilon_{n_i}\right).
\label{eq:ZN_canonical}
\end{equation}
We start by rewriting the canonical sum using occupation variables. For that, 
let us introduce occupation numbers $\eta_n\in\{0,1\}$ with the fixed-$N$ constraint 
\begin{equation}
\sum_{n=0}^\infty \eta_n = N,
\qquad
E[\eta]=\sum_{n=0}^\infty \eta_n \varepsilon_n .
\end{equation}
Then the canonical partition function becomes
\begin{equation}
Z_N = \sum_{\{\eta\}} e^{-\beta E[\eta]}\,\delta_{\sum_n\eta_n,\,N},
\label{eq:ZN_eta}
\end{equation}
and the sum over occupied levels can be rewritten as
\[
\sum_{j\neq k} (\cdots)
=
\sum_{m\neq n}\eta_m\eta_n\,(\cdots)_{m,n},
\]
where $(\cdots)_{m,n}$ means: replace $(n_j,n_k)$ by $(m,n)$.
We define the symmetric kernel-like building block
\begin{equation}
A_{mn}(x)
:=
\left[\phi_m(x)\phi_n'(x)\right]^2
-\phi_m(x)\phi_n(x)\phi_m'(x)\phi_n'(x).
\label{eq:Amn_def}
\end{equation}
Then \eqref{eq:FN_original} becomes
\begin{equation}
F_N(x)
=
\sum_{m\neq n} \left\langle \eta_m\eta_n \right\rangle_N\,A_{mn}(x),
\label{eq:FN_eta_avg}
\end{equation}
with the canonical expectation
\begin{equation}
\langle \mathcal O(\eta)\rangle_N
:=
\frac{1}{Z_N}\sum_{\{\eta\}} e^{-\beta E[\eta]}\,
\delta_{\sum_n\eta_n,\,N}\,
\mathcal O(\eta).
\label{eq:canonical_avg_def}
\end{equation}
Note that we could also include the $m=n$ terms in \eqref{eq:FN_eta_avg}
since $A_{nn}(x)=0$.

The grand partition function $\Xi(z)$ is defined in
\eqref{eq:Xi_def_main}, with the canonical $Z_N$ recovered through
the contour representation \eqref{eq:ZN_contour}. Similarly, for any
function of occupations $\mathcal O(\eta)$,
\begin{equation}
\sum_{\{\eta\}} e^{-\beta E[\eta]}\,\delta_{\sum\eta,N}\,\mathcal O(\eta)
=
\frac{1}{2\pi i}\oint_{\mathcal C}\frac{dz}{z^{N+1}}\;
\sum_{\{\eta\}} e^{-\beta E[\eta]} z^{\sum_n\eta_n}\,\mathcal O(\eta).
\label{eq:O_contour_general}
\end{equation}

Now, for fixed $z$, the occupations are independent Bernoulli random variables with
\[
\mathbb P_z(\eta_n=1)=f_n(z),
\qquad
f_n(z):=\frac{z e^{-\beta\varepsilon_n}}{1+z e^{-\beta\varepsilon_n}}.
\]
Equivalently, for fixed $z$,
\begin{equation}
\sum_{\{\eta\}} e^{-\beta E[\eta]} z^{\sum\eta}
=
\Xi(z),
\qquad
\sum_{\{\eta\}} e^{-\beta E[\eta]} z^{\sum\eta}\,\eta_m\eta_n
=
\Xi(z)\,f_m(z)f_n(z)
\quad (m\neq n).
\end{equation}
Therefore, for $m\neq n$,
\begin{equation}
\langle \eta_m\eta_n\rangle_N
=
\frac{1}{Z_N}\frac{1}{2\pi i}\oint_{\mathcal C}\frac{dz}{z^{N+1}}\;
\Xi(z)\,f_m(z)f_n(z).
\label{eq:etaeta_contour}
\end{equation}
Plugging \eqref{eq:etaeta_contour} into \eqref{eq:FN_eta_avg} yields an exact contour representation:
\begin{align}
F_N(x)
&=
\sum_{m\neq n} A_{mn}(x)\,
\frac{1}{Z_N}\frac{1}{2\pi i}\oint_{\mathcal C}\frac{dz}{z^{N+1}}\;
\Xi(z)\,f_m(z)f_n(z)
\nonumber\\
&=
\frac{1}{Z_N}\frac{1}{2\pi i}\oint_{\mathcal C}\frac{dz}{z^{N+1}}\;
\Xi(z)\,
\sum_{m\neq n} f_m(z)f_n(z)\,A_{mn}(x).
\label{eq:FN_contour_sum}
\end{align}

The final step is to define the $z$-kernel
\begin{equation}
K_z(x,y):=\sum_{n=0}^\infty f_n(z)\,\phi_n(x)\phi_n(y),
\label{eq:Kz_def}
\end{equation}
together with its associated local quantities
\begin{align}
\rho_z(x) &:= K_z(x,x)=\sum_{n\ge 0} f_n(z)\,\phi_n(x)^2,
\\
S_z(x) &:= \partial_y K_z(x,y)\big|_{y=x}=\sum_{n\ge 0} f_n(z)\,\phi_n(x)\phi_n'(x),
\\
\kappa_z(x) &:= \partial_x\partial_y K_z(x,y)\big|_{y=x}=\sum_{n\ge 0} f_n(z)\,\phi_n'(x)^2.
\end{align}
Expanding $\rho_z\kappa_z-S_z^2$: 
\begin{align}
\rho_z(x)\kappa_z(x)-S_z(x)^2
&=
\sum_{m,n\ge 0} f_m(z)f_n(z)
\left[
\phi_m(x)^2\phi_n'(x)^2-\phi_m(x)\phi_m'(x)\phi_n(x)\phi_n'(x)
\right]
\nonumber\\
&=
\sum_{m,n\ge 0} f_m(z)f_n(z)\,A_{mn}(x).
\label{eq:Fz_expand}
\end{align}
But $A_{nn}(x)=0$, so
\begin{equation}
\sum_{m\neq n} f_m(z)f_n(z)\,A_{mn}(x)
=
\rho_z(x)\kappa_z(x)-S_z(x)^2.
\label{eq:offdiag_equals_full}
\end{equation}
Therefore, the inner sum in \eqref{eq:FN_contour_sum} collapses to a kernel functional:
\begin{equation}
F_N(x)
=
\frac{1}{Z_N}\frac{1}{2\pi i}\oint_{\mathcal C}\frac{dz}{z^{N+1}}\;
\Xi(z)\,
\left[\rho_z(x)\kappa_z(x)-S_z(x)^2\right].
\label{eq:FN_kernel_contour_final}
\end{equation}
Equivalently, entirely in terms of $K_z$,
\begin{equation}
F_N(x)
=
\frac{1}{Z_N}\frac{1}{2\pi i}\oint_{\mathcal C}\frac{dz}{z^{N+1}}\;
\Xi(z)\,
\left[
K_z(x,x)\,\partial_x\partial_y K_z(x,y)\big|_{y=x}
-
\left(\partial_y K_z(x,y)\big|_{y=x}\right)^2
\right].
\label{eq:FN_kernel_contour_onlyK}
\end{equation}

%=============================================================
\subsection{Pair representation}
\label{sec:kernel}
%=============================================================
For a fixed Slater determinant built from orbitals
$\{n_1,\ldots,n_N\}$, define the configuration kernel
\begin{equation}
  K_{\mathrm{config}}(x,y)
  = \sum_{a:\,n_a=1} \phi_a(x)\,\phi_a(y),
\end{equation}
and the local quantities
\begin{equation}
  \rho = K(x,x),
  \qquad
  S = \partial_y K(x,y)\big|_{y=x},
  \qquad
  \kappa = \partial_x\partial_y K(x,y)\big|_{y=x}.
\end{equation}

For any set of occupation numbers $\{n_a\}\in\{0,1\}$,
\begin{equation}\label{eq:pair-sum}
  \sum_{j\neq k} A_{n_j,n_k}
  = \rho_{\mathrm{config}}\,\kappa_{\mathrm{config}}
    - S_{\mathrm{config}}^2,
\end{equation}
where $A_{a,b} = (\phi_a\,\phi_b')^2
- \phi_a\,\phi_b\,\phi_a'\,\phi_b'$.
Expanding the right-hand side we have 
\begin{align}
  \rho\,\kappa - S^2
  &= \left(\sum_a n_a\,\phi_a^2\right)
     \left(\sum_b n_b\,(\phi_b')^2\right)
   - \left(\sum_a n_a\,\phi_a\,\phi_a'\right)^2
  \\
  &= \sum_{a,b} n_a\,n_b
     \left[\phi_a^2\,(\phi_b')^2
     - \phi_a\,\phi_a'\,\phi_b\,\phi_b'\right]
  \\
  &= \sum_{a,b} n_a\,n_b\,A_{a,b}(x).
\end{align}
Since $A_{a,a} = \phi_a^2\,(\phi_a')^2
- \phi_a^2\,(\phi_a')^2 = 0$, the diagonal terms
vanish and the double sum reduces to
$\sum_{a\neq b} n_a\,n_b\,A_{a,b}$.

Symmetrising, $A_{a,b} + A_{b,a}
= (\phi_a\,\phi_b' - \phi_b\,\phi_a')^2$, so the
contribution from each unordered pair $(a,b)$ with $a<b$ is
the squared Wronskian.  Defining the pair integral
\begin{equation}\label{eq:Jab}
  J_{ab}
  = \int dx\,
    \left(\phi_a\,\phi_b' - \phi_b\,\phi_a'\right)^2.
\end{equation}

Averaging~\eqref{eq:pair-sum} over the canonical ensemble,
\begin{equation}\label{eq:contact-pairs}
  \mathcal C_N^\CE
  = \frac{2}{\pi}\sum_{a<b}
    \langle n_a\,n_b\rangle_\CE\,J_{ab}.
\end{equation}

If levels are statistically independent
(as in the GCE), then for $a\neq b$,
$\langle n_a\,n_b\rangle_\GCE = \nbar_a\,\nbar_b$
where $\nbar_a = (1+e^{\beta(\varepsilon_a-\mu)})^{-1}$.
The contact takes the kernel form
\begin{equation}
  \mathcal C_N^\GCE
  = \frac{1}{\pi}\,\bar{\vb{n}}^T J\,\bar{\vb{n}}.
\end{equation}

The fixed-$N$ constraint introduces correlations:
\begin{equation}
  \langle n_a\,n_b\rangle_\CE
  = \nbar_a^\CE\,\nbar_b^\CE
    + \mathrm{Cov}_\CE(n_a,n_b).
\end{equation}
The covariance is negative (anti-bunching from the
particle-number constraint) and scales as
$\mathrm{Cov}_\CE(n_a,n_b)
\sim -\nbar_a(1-\nbar_a)\,\nbar_b(1-\nbar_b)
/\mathrm{Var}_\GCE(N)$.
At finite~$\tau$, the total ensemble difference
$\Delta \mathcal C := \mathcal C_N^\CE - \mathcal C_N^\GCE$ scales as
$\sim N^{3/2}$, and therefore  {cannot} be neglected
at subleading order.

An exact canonical computation that avoids enumerating
all $\binom{M}{N}$ configurations uses the fugacity
contour integral.  Define the grand partition function
$\Xi(z) = \prod_{n=0}^{M-1}(1+z\,q_n)$ with
$q_n = e^{-\beta\varepsilon_n}$, and the fugacity-dependent
occupation $\nbar_a(z) = z\,q_a/(1+z\,q_a)$.  Then
\begin{equation}\label{eq:contour}
  \mathcal C_N^\CE
  = \frac{2}{\pi}\,\frac{1}{Z_N}\,
    \frac{1}{2\pi i}\oint\frac{dz}{z^{N+1}}\,
    \Xi(z)\,G(z),
\end{equation}
where
\begin{equation}
  G(z) = \sum_{a,b}\nbar_a(z)\,\nbar_b(z)\,R_{ab},
  \qquad
  R_{ab} = \int dx\,
  \left[\phi_a^2\,(\phi_b')^2
  - \phi_a\,\phi_a'\,\phi_b\,\phi_b'\right].
\end{equation}
Note that $J_{ab} = R_{ab} + R_{ba}$.

At fixed complex fugacity~$z$, the occupation numbers are
independent (the kernel form is exact), so the integrand
is well-defined.  The contour integral projects onto the
$N$-particle sector.

%%%%%%%%%%%%%%%%%%%%%% ASYMPTOTIC EXPANSIONS %%%%%%%%%%%%%%%%%%%%%%%%
\section{Asymptotic expansions of the scaling functions}
\label{app:asymptotics}

This appendix collects the derivations of the closed asymptotic
limits of $A(\tau)$ and $B(\tau)$ quoted in the main text:
the Sommerfeld expansion \eqref{eq:A_tau_lowtau} and the virial
expansion \eqref{eq:A_hightau_final} of the leading coefficient,
and the Sommerfeld result \eqref{eq:Bsad_lowtau_final_closed}
 and the virial expansion
\eqref{eq:Bsad_hightau_final_1overTau} of the subleading
coefficient.

% ===============================================================
\subsection{Sommerfeld expansion of $A(\tau)$}
\label{app:asymptotics_A_low}

For each fixed $\tau\ge0$ the integrals defining $A(\tau)$ are
finite. In the low-$\tau$ expansion we ignore exponentially small
corrections to $\xi(\tau)$, recall
$\xi(\tau\ll1)=1-\tau e^{-1/\tau}+\mathcal O(e^{-2/\tau})$ from
\eqref{eq:xi_eq}, and replace $\xi$ by its zero-temperature
value $\xi(0)=1$ throughout the algebraic Sommerfeld series; the
neglected pieces enter the contact only at
$\mathcal O(e^{-1/\tau})$ and are invisible in the polynomial
expansion. The Fermi factor \eqref{eq:f_tau} depends on $q$ only
through $q^{2}-a$, with $a:=1-u^{2}$. For a smooth function
$g(\varepsilon)$ one has the standard Sommerfeld expansion
\begin{equation}
\int_0^\infty d\varepsilon\; g(\varepsilon)\,
\frac{1}{e^{(\varepsilon-a)/\tau}+1}
=
\int_0^a d\varepsilon\; g(\varepsilon)
+\frac{\pi^2\tau^2}{6}\,g'(a)
+\mathcal O(\tau^4),
\qquad (a>0),
\label{eq:sommerfeld_general}
\end{equation}
which yields, for $|u|<1$ (i.e.\ $a>0$),
\begin{align}
I_0(u;\tau)\,I_2(u;\tau)
&=
\left(2\sqrt{a}-\frac{\pi^2\tau^2}{12}\,a^{-3/2}\right)
\left(\frac{2}{3}a^{3/2}+\frac{\pi^2\tau^2}{12}\,a^{-1/2}\right)
\nonumber\\
&=\frac{4}{3}(1-u^2)^2+\frac{\pi^2}{9}\,\tau^2.
\label{eq:I0I2_expanded}
\end{align}
Remarkably, the $\tau^2$ term is independent of $u$ in the bulk
region $|u|<1$. For $|u|>1$ the quantity $a=1-u^2<0$ and both
$I_0$ and $I_2$ are exponentially small in $1/\tau$; moreover, the
turning-point region $|u|\approx 1$ has width $\Delta u\sim \tau$
and contributes only $\mathcal O(\tau^3)$ to the $u$ integral,
hence it does not affect the $\tau^2$ coefficient. Finally, we
find
\begin{align}
\int_{-\infty}^{\infty}du\; I_0(u;\tau)I_2(u;\tau)
&=
\frac{4}{3}\int_{-1}^{1}du\,(1-u^2)^2
+\frac{\pi^2}{9}\tau^2\int_{-1}^{1}du
\nonumber\\
&=
\frac{64}{45}+\frac{2\pi^2}{9}\tau^2,
\label{eq:u_integral_result}
\end{align}
which, inserted into \eqref{eq:A_tau_def}, gives
Eq.~\eqref{eq:A_tau_lowtau} of the main text.

% ===============================================================
\subsection{Virial expansion of $A(\tau)$}
\label{app:asymptotics_A_high}

For $\tau\gg 1$ the saddle fugacity is small and the gas is in the
Boltzmann regime. Writing the Fermi factor \eqref{eq:f_tau} as
\begin{equation}
f_\tau(q,u)=\frac{1}{z^{-1}e^{(q^2+u^2)/\tau}+1},
\qquad z:=e^{\xi/\tau},
\end{equation}
the LDA number constraint \eqref{eq:xi_eq} reads $\tau\log(1+z)=1$,
hence $z=e^{1/\tau}-1=1/\tau+\mathcal O(1/\tau^2)\ll 1$ at high $\tau$.
Using the virial (small-$z$) series
\begin{equation}
f_\tau(q,u)= z\,e^{-(q^2+u^2)/\tau}-z^2 e^{-2(q^2+u^2)/\tau}+\mathcal O(z^3),
\label{eq:f_virial_A}
\end{equation}
the moments \eqref{eq:I0_I2_def} reduce to elementary Gaussian
integrals. Writing
\begin{equation}
I_0(u;\tau)=z\,G_1(u;\tau)-z^2 G_2(u;\tau)+\mathcal O(z^3),
\quad
I_2(u;\tau)=z\,H_1(u;\tau)-z^2 H_2(u;\tau)+\mathcal O(z^3),
\label{eq:GH_virial_def}
\end{equation}
the Gaussian moments read
\begin{equation}
G_1=\sqrt{\pi\tau}\,e^{-u^2/\tau},\quad
G_2=\sqrt{\pi\tau/2}\,e^{-2u^2/\tau},\quad
H_1=\frac{\sqrt\pi}{2}\tau^{3/2}e^{-u^2/\tau},\quad
H_2=\frac{\sqrt\pi}{2}\tau^{3/2}\,2^{-3/2}e^{-2u^2/\tau}.
\label{eq:GH_virial_moments_A}
\end{equation}
Hence
\begin{align}
g_0(u;\tau) = I_0(u;\tau) I_2(u;\tau)
&= z^2\,G_1 H_1 - z^3\,(G_1 H_2+G_2 H_1)+\mathcal O(z^4) \nonumber\\
&= \frac{\pi}{2}\,\tau^2 z^2 e^{-2u^2/\tau}
   -\frac{\pi}{2}\,\tau^2 z^3\,\frac{3}{2\sqrt 2}\,e^{-3u^2/\tau}+\mathcal O(z^4).
\label{eq:g0_hightau}
\end{align}
Integrating over $u$ we have
\begin{equation}
\int du\;g_0(u;\tau)
=
\frac{\pi^{3/2}}{2\sqrt 2}\,\tau^{5/2}z^2
\left(
1-\frac{\sqrt3}{2}\,z
\right)+\mathcal O(z^4).
\label{eq:intg0_hightau}
\end{equation}
Plugging \eqref{eq:intg0_hightau} into \eqref{eq:A_tau_def} yields
\begin{equation}
A(\tau)
=
\frac{\tau^{5/2}}{\pi^{3/2}}\,z^2
\left(1-\frac{\sqrt3}{2}\,z\right).
\label{eq:A_in_z}
\end{equation}
At $\tau\gg 1$, the number constraint in scaled variables gives
\begin{equation}
\frac{1}{\tau}=z-\frac{1}{2}z^2+\mathcal O(z^3)
\qquad\Longrightarrow\qquad
z=\frac{1}{\tau}+\frac{1}{2\tau^2}+ \mathcal O \left(\frac{1}{\tau^3}\right).
\label{eq:z_vs_tau_A}
\end{equation}
Inserting \eqref{eq:z_vs_tau_A} into \eqref{eq:A_in_z}:
\begin{equation}
\tau^{5/2}z^2
=
\sqrt{\tau}\left(1+\frac{1}{\tau}+\mathcal O \left(\tau^{-2}\right)\right),
\qquad
1-\frac{\sqrt3}{2}z
=
1-\frac{\sqrt3}{2\tau}+\mathcal O \left(\tau^{-2}\right);
\end{equation}
multiplying the two expansions to relative order $1/\tau$ gives
Eq.~\eqref{eq:A_hightau_final} of the main text.

% ===============================================================
\subsection{Sommerfeld expansion of $B(\tau)$: bulk contribution}
\label{app:asymptotics_B_low}

In the bulk region $|u|<1$, with $a:=1-u^2>0$, the moments
\eqref{eq:I0_I2_def} admit the standard Sommerfeld expansion
\begin{equation}
I_0(u;\tau)=2\sqrt{a}-\frac{\pi^2\tau^2}{12}\,a^{-3/2}+ \mathcal O(\tau^4),
\qquad
I_2(u;\tau)=\frac{2}{3}a^{3/2}+\frac{\pi^2\tau^2}{12}\,a^{-1/2}+\mathcal O(\tau^4).
\label{eq:I0I2_lowtau}
\end{equation}
Using the identity
$f_\tau(1-f_\tau)=\tau\,[-\partial_\varepsilon f_\tau]$ with
$\varepsilon=q^2$ in \eqref{eq:J0_J2}, together with the
Sommerfeld formula
\begin{equation}
\int_0^\infty d\varepsilon\, g(\varepsilon)\,f_\tau(1-f_\tau)
=
\tau \left[g(a)+\frac{\pi^2\tau^2}{6}\,g''(a)+ \mathcal O(\tau^4)\right],
\qquad (a>0),
\label{eq:sommerfeld_for_J}
\end{equation}
and the analogous identity
$f_\tau(1-f_\tau)(1-2f_\tau)=\tau^2\,\partial_\varepsilon^2 f_\tau$ for
\eqref{eq:K0_K2}, we obtain
\begin{equation}
J_0=\tau\,a^{-1/2} + \mathcal O \left(\tau ^3 \right),
\quad
J_2=\tau\,a^{1/2}  + \mathcal O \left(\tau ^3 \right),
\label{eq:J0J2_lowtau}
\end{equation}
\begin{equation}
K_0= -\frac{1}{2}\,\tau^2\,a^{-3/2} +  \mathcal O \left(\tau ^4 \right),
\quad
K_2=  \frac{1}{2}\,\tau^2\,a^{-1/2} +  \mathcal O \left(\tau ^4 \right),
\label{eq:K0K2_lowtau}
\end{equation}
valid for $|u|<1$ away from the turning points. Substituting
\eqref{eq:I0I2_lowtau}--\eqref{eq:K0K2_lowtau} into
\eqref{eq:g0_t}--\eqref{eq:g0_tt}, the leading-order coefficients
combine into
\begin{equation}
\partial_t g_0(u;\tau)=\frac{8}{3}\,\tau\,a +  \mathcal O \left(\tau ^3 \right),
\quad
\partial_t^2 g_0(u;\tau)=\frac{8}{3}\,\tau^2 +  \mathcal O \left(\tau ^4 \right),
\label{eq:dtg_lowtau}
\end{equation}
where, remarkably, the $\tau^2$ term in $\partial_t^2 g_0$ is
$u$-independent in the bulk and the would-be $\tau^4$ correction
cancels in the combination $K_0I_2+2J_0J_2+I_0K_2$; this
establishes Eq.~\eqref{eq:dtg2_bulk_main} of the main text.

The global functions \eqref{eq:kappa2_scaling}--\eqref{eq:kappa3_scaling}
are most cleanly evaluated as two-dimensional phase-space integrals.
Setting $u=r\cos\theta$, $q=r\sin\theta$, the Fermi factor depends
only on $r^2$; substituting $s=r^2$ converts
$V_2,V_3$ into one-dimensional integrals over the Fermi function
$f(y)=(e^y+1)^{-1}$ of argument $(s-\xi)/\tau$. Using
$f'(y)=-f(y)[1-f(y)]$ and
$f(1-f)(1-2f)=-\frac{d}{dy}[f(1-f)]$, the integrals telescope into
the identities quoted in
Eq.~\eqref{eq:V2V3_exact_main}; with
$\xi=1+\mathcal O(e^{-1/\tau})$ these give
\begin{equation}
V_2(\tau)=\pi\tau\,f(-1/\tau)
=\pi\tau\left[1+ \mathcal O \left( e^{-1/\tau} \right)\right],
\label{eq:V2_lowtau_exact}
\end{equation}
\begin{equation}
V_3(\tau)=\pi\tau\,f(-1/\tau) \left[1-f(-1/\tau)\right]
=\pi\tau\,e^{-1/\tau}\left[1+ \mathcal O \left( e^{-1/\tau} \right)\right].
\label{eq:V3_lowtau_exact}
\end{equation}
Thus $V_3(\tau)$ is nonperturbatively small: it vanishes to all
algebraic orders in the Sommerfeld expansion. Consequently
$V_3/V_2^{\,2} \sim \mathcal O(e^{-1/\tau}/\tau)$, and the $V_3$-term in
\eqref{eq:H_def} contributes only $\mathcal O(e^{-1/\tau})$ to $B(\tau)$; its
algebraic expansion is therefore controlled solely by
$\partial_t^2 g_0/V_2$.

Combining \eqref{eq:dtg_lowtau} and \eqref{eq:V2_lowtau_exact} in
\eqref{eq:H_def} gives, for $|u|<1$,
\begin{equation}
\mathcal H(u;\tau)
=
\frac{4}{3\pi}\,\tau+ \mathcal O \left(\tau^3 \right)+\mathcal O \left(e^{-1/\tau}\right),
\label{eq:H_lowtau_leading}
\end{equation}
which is $u$-independent in the bulk. Inserting
\eqref{eq:H_lowtau_leading} into \eqref{eq:Bsad_master} and
integrating over $u\in[-1,1]$ yields the bulk contribution
\begin{equation}
B(\tau)=-\frac{16\sqrt2}{3\pi^3}\,\tau\;+\;\text{(edge)}\;+\;\mathcal O \left(\tau^3 \right) +
\mathcal O \left(e^{-1/\tau}\right),
\label{eq:Bsad_linear_plus_edge}
\end{equation}
where ``(edge)'' denotes the boundary-layer contributions analysed
in Appendix~\ref{app:asymptotics_B_edge}, which are shown there to
vanish at order $\tau^{2}$.

% ===============================================================
\subsection{Edge layers of $B(\tau)$}
\label{app:asymptotics_B_edge}

The bulk Sommerfeld formulae \eqref{eq:K0K2_lowtau} are not uniform
at the turning points $u=\pm 1$, where $a\to 0$: the negative powers
of $a$ diverge, signalling boundary layers of width $1-u^2\sim\tau$
that must be resolved separately. By the $u\to-u$ symmetry it
suffices to analyse the layer at $u=+1$ and double the result. We
introduce the edge scaling variable
\begin{equation}
y:=\frac{1-u^2}{\tau},\qquad
du=-\frac{\tau\,dy}{2\sqrt{1-\tau y}}
\;\;\xrightarrow[\tau\to 0]{}\;\;-\frac{\tau}{2}\,dy ,
\label{eq:y_def_edge}
\end{equation}
with $y>0$ inside the cloud ($|u|<1$) and $y<0$ on the classically
forbidden side ($|u|>1$), and rescale momenta by $q=\sqrt{\tau}\,p$,
so that
$q^2+u^2-\xi(\tau) = \tau(p^2 - y) + \mathcal O(\tau\,e^{-1/\tau})$ when
$\xi(\tau) = 1 + \mathcal O(e^{-1/\tau})$. The Fermi factor reduces
to the universal, $\tau$-independent form
\begin{equation}
f_\tau(\sqrt{\tau}\,p,\,u)\xrightarrow[\tau\to 0]{}
f_y(p):=\frac{1}{e^{p^2-y}+1},
\label{eq:f_y_edge}
\end{equation}
whose moments $\Phi_0(y)$ and $\Phi_2(y)$ were defined in
Eq.~\eqref{eq:Phi0Phi2_def} of the main text. Using the identity
$\partial_t = z\,\partial_z = \tau\,\partial_\xi$ at
fixed $u$, which in edge variables becomes $\partial_t \to \partial_y$
(a shift in $\xi$ by $\delta\xi$ shifts $y$ by $\delta\xi/\tau$, and
$\partial_t = \tau\,\partial_\xi$, so the two factors of $\tau$ cancel),
combined with $dq = \sqrt{\tau}\,dp$, the local integrals acquire
the systematic edge scaling
\begin{align}
I_0 &= \sqrt{\tau}\,\Phi_0(y),
\quad
I_2 = \tau^{3/2}\,\Phi_2(y),
\nonumber\\
J_0 &= \sqrt{\tau}\,\Phi_0'(y),
\quad
J_2 = \tau^{3/2}\,\Phi_2'(y),
\nonumber \\
K_0 &= \sqrt{\tau}\,\Phi_0''(y),
\quad
K_2 = \tau^{3/2}\,\Phi_2''(y).
\label{eq:edge_scalings}
\end{align}
Substituting \eqref{eq:edge_scalings} into the combination appearing
in $\partial_t^2 g_0$ gives
\begin{equation}
K_0 I_2+2J_0 J_2+I_0 K_2
=
\tau^2\left[\Phi_0''\Phi_2+2\Phi_0'\Phi_2'+\Phi_0\Phi_2''\right]
=
\tau^2\,\mathcal P''(y),
\quad
\mathcal P(y):=\Phi_0(y)\Phi_2(y),
\label{eq:combo_is_second_derivative}
\end{equation}
i.e.\ the combination collapses to the second derivative of a single
universal function. Since $V_2$ is uniform in $u$ at this order,
the $u$-integration of $\partial_t^2 g_0/V_2$ collects the edge
correction.

We compute the $u$-integral by matched asymptotic subtraction,
taking care to include both sides of each layer: the bulk
formula \eqref{eq:dtg_lowtau} holds for $y\gg1$ (deep inside the
cloud), while on the outer side $y<0$ the integrand, though
exponentially small for $|y|\gg1$, is of order $\tau^{2}$ within the
layer $|y|=\mathcal O(1)$ and cannot be discarded. Splitting the
full line symmetrically and substituting \eqref{eq:y_def_edge} near
$u=+1$,
\begin{align}
\int_{-\infty}^{\infty} du\,\partial_t^2 g_0(u;\tau)
&= 2\int_0^1 du\,\frac{8\,\tau^2}{3}
\;+\;
2\int_0^{\infty} du\,\left[\partial_t^2 g_0(u;\tau)
- \frac{8\,\tau^2}{3}\,\Theta(1-u)\right]
\nonumber\\
&= \frac{16\,\tau^2}{3}
\;+\;
\tau^3\int_{-\infty}^{\infty} dy\,\left[\mathcal P''(y) - \frac{8}{3}\,\Theta(y)\right]
\;+\;\mathcal O(\tau^4),
\label{eq:matched_asymptotic_subtraction}
\end{align}
where $\Theta$ is the Heaviside function: the $y>0$ part of the
subtraction integral is the inner-layer contribution and the $y<0$
part the outer-layer one.

Both parts are total derivatives and are fixed by the asymptotics
of $\mathcal P'$. As $y\to\infty$, the Sommerfeld expansion of the
edge moments gives
\begin{equation}
\Phi_0(y)=2\sqrt y-\frac{\pi^2}{12}\,y^{-3/2}+\mathcal O(y^{-7/2}),
\qquad
\Phi_2(y)=\frac23\,y^{3/2}+\frac{\pi^2}{12}\,y^{-1/2}+\mathcal O(y^{-5/2}),
\label{eq:edge_large_y}
\end{equation}
so that
\begin{equation}
\mathcal P(y)=\frac43\,y^2+\frac{\pi^2}{9}+\mathcal O(y^{-2}),
\qquad
\mathcal P'(y)=\frac83\,y+\mathcal O(y^{-3}),
\label{eq:P_large_y}
\end{equation}
the constant $\pi^{2}/9$ matching the uniform bulk
Sommerfeld correction in \eqref{eq:I0I2_expanded}, a closure check
of the asymptotics. As $y\to-\infty$, $\mathcal P$ and
$\mathcal P'$ vanish as $e^{2y}$. Therefore
\begin{align}
\int_{0}^{\infty} dy\,\left[\mathcal P''(y)-\frac83\right]
&=\left[\mathcal P'(y)-\frac83\,y\right]_{0}^{\infty}
=-\,\mathcal P'(0),
\nonumber\\
\int_{-\infty}^{0} dy\,\mathcal P''(y)
&=\mathcal P'(0)-\mathcal P'(-\infty)=+\,\mathcal P'(0):
\label{eq:inner_outer_split}
\end{align}
the inner and outer contributions are equal and opposite, and the
subtraction integral in \eqref{eq:matched_asymptotic_subtraction}
vanishes, 
\begin{equation}
\int_{-\infty}^{\infty} dy\,\left[\mathcal P''(y) - \frac{8}{3}\,\Theta(y)\right]=0 .
\label{eq:edge_cancellation}
\end{equation}
The $\mathcal O(\tau^{2})$ edge contribution to $B(\tau)$
therefore cancels between the two sides of the Fermi surface.
Combining
\eqref{eq:matched_asymptotic_subtraction}--\eqref{eq:edge_cancellation}
with \eqref{eq:V2_lowtau_exact} in \eqref{eq:Bsad_master}
establishes the low-temperature law
\eqref{eq:Bsad_lowtau_final_closed} of the main text.

% ===============================================================
\subsection{Virial expansion of $B(\tau)$}
\label{app:asymptotics_B_high}

The first $1/\tau$ correction to the Boltzmann limit
\eqref{eq:Bsad_hightau_leading} follows from extending the virial
expansion \eqref{eq:f_virial_A} of $f_\tau$ to next order. Recalling the
virial decomposition \eqref{eq:GH_virial_def} and the Gaussian moments
\eqref{eq:GH_virial_moments_A}, the high-$\tau$ form of $g_0$ is given by
\eqref{eq:g0_hightau}, which we rewrite in the compact form
\begin{equation}
g_0(u;\tau)
=z^2\,\frac{\pi}{2}\tau^2 e^{-2u^2/\tau}
 \left[1-z\,\frac{3}{2^{3/2}}\,e^{-u^2/\tau}+\mathcal O(z^2)\right].
\label{eq:g0_virial}
\end{equation}
The virial expansion of $f_\tau(1-f_\tau)=f_\tau-f_\tau^2+\cdots$
gives, in turn,
\begin{equation}
J_\alpha=I_\alpha-z\,\Delta_\alpha+\mathcal O(z^3),
\qquad
K_\alpha=I_\alpha-3z\,\Delta_\alpha+\mathcal O(z^3),
\label{eq:JK_corrections}
\end{equation}
where $\Delta_\alpha$ denotes the $\mathcal O(z^2)$ piece of $I_\alpha$
(i.e.\ $\Delta_0 = G_2$, $\Delta_2 = H_2$). The
corresponding corrections to \eqref{eq:dtg_boltz} read
\begin{align}
\partial_t g_0   &= 2 g_0 - z^3\,Y(u;\tau) + \mathcal O(z^4),
\label{eq:dtg0_NLO}\\
\partial_t^2 g_0 &= 4 g_0 - 5\,z^3\,Y(u;\tau) + \mathcal O(z^4),
\label{eq:dttg0_NLO}
\end{align}
with $Y(u;\tau):= G_1 H_2+G_2 H_1
=(\pi/2)\,\tau^{2}\,(3/2^{3/2})\,e^{-3u^2/\tau}$. Likewise, integrating
\eqref{eq:JK_corrections} over $u$,
\begin{equation}
V_2(\tau)=\pi\tau\,\left[z-z^2+\mathcal O\left(z^3\right)\right],
\qquad
V_3(\tau)=\pi\tau\,\left[z-2z^2+\mathcal O \left(z^3\right)\right],
\label{eq:V2V3_virial}
\end{equation}
so that $V_3/V_2^{\,2}=(\pi\tau\,z)^{-1}[1+\mathcal O(z^2)]$ has  {no}
$\mathcal O(z)$ correction (the corrections to $V_3$ and $V_2^{\,2}$ cancel at
this order). Inserting
\eqref{eq:dtg0_NLO}--\eqref{eq:V2V3_virial} into \eqref{eq:H_def} and
integrating, then using the number constraint
\eqref{eq:z_vs_tau_A} [$z=1/\tau+1/(2\tau^2)+\mathcal O(\tau^{-3})$],
one obtains Eq.~\eqref{eq:Bsad_hightau_final_1overTau} of the main
text.

\section{Absence of competing corrections at order $N^{3/2}$}
\label{app:powercounting}

The subleading coefficient \eqref{eq:Bsad_oneline} was derived from
a single mechanism: the Gaussian fluctuation of the fugacity around
the canonical saddle. Because the leading contact density is
locally of order $N^{2}$ and is integrated over a region of size
$\mathcal O(\sqrt N)$, any correction of relative order
$1/N$ to the local integrand would contribute at the same $N^{3/2}$
order. This appendix enumerates the possible sources of such
corrections and shows that each enters only at relative order
$1/N^{2}$, i.e.\ at absolute order $N^{1/2}$ in the contact.
Throughout we use the fixed-$\tau$ scaling variables
\eqref{eq:scaling_xp}, in which $\beta=1/(\tau N)$,
$\mu=\xi(\tau)N$, and each spatial gradient of a scaled quantity
carries a factor $N^{-1/2}$.

The numerator of \eqref{eq:FN_contour} carries the extra factor
$\widetilde G(t;x)$, so its effective action
$\psi(t)+\log\widetilde G(t)-Nt$ is stationary not at $t_\star$ but
at the displaced point
\begin{equation}
t_\star+\delta t,
\qquad
\delta t=-\frac{\widetilde G_1}{\widetilde G_0\,\Phi_2}
=\mathcal O\left(\frac1N\right),
\label{eq:delta_t_shift}
\end{equation}
since $\widetilde G_1/\widetilde G_0=\mathcal O(1)$ and
$\Phi_2=\mathcal O(N)$. To see that expanding both integrals about
the common point $t_\star$ loses nothing, expand instead
about an arbitrary nearby point $t_0=t_\star+\epsilon$ with
$\epsilon=\mathcal O(1/N)$ [in particular $t_0$ may be the
displaced saddle \eqref{eq:delta_t_shift}]. Since
$\Phi'(t_0)=\Phi_2\,\epsilon+\mathcal O(N\epsilon^{2})$ no longer
vanishes, the Laplace ratio formula acquires one additional term at
the working order,
\begin{equation}
F_N=\widetilde G(t_0)
-\frac{\Phi'(t_0)}{\Phi_2}\,\widetilde G_1
-\frac{\widetilde G_2}{2\Phi_2}
+\frac{\Phi_3}{2\Phi_2^{2}}\,\widetilde G_1
+\mathcal O(N^{-2})\,\widetilde G_0 ,
\label{eq:offsaddle_ratio}
\end{equation}
and the first two terms combine to
$\widetilde G_0+\widetilde G_1\epsilon-\widetilde G_1\epsilon
+\mathcal O(N^{-2})\widetilde G_0$: the $\epsilon$-dependence
cancels. The ratio expansion
\eqref{eq:ratio_expansion_t} is therefore invariant under
$\mathcal O(1/N)$ changes of the common expansion point. Expanding
about $t_\star$, where $\Phi'=0$, is the most economical choice, and the displacement of the numerator saddle requires no
separate treatment: its effect at order $N^{3/2}$ is the
$\widetilde G_1$ term of \eqref{eq:ratio_expansion_t}.

The saddle condition \eqref{eq:saddle_number} is a discrete sum,
whereas the scaled number equation \eqref{eq:xi_eq} is its
continuum limit. For the oscillator spectrum
$\varepsilon_n=n+\frac12$ the sum is of midpoint type: for
a smooth function $h$ decaying at infinity, the midpoint
Euler--Maclaurin formula reads
\begin{equation}
\sum_{n=0}^{\infty}h\left(n+\frac12\right)
=\int_0^\infty h(x)\,dx+\frac{1}{24}\,h'(0)
-\frac{7}{5760}\,h'''(0)+\cdots,
\label{eq:euler_maclaurin_midpoint}
\end{equation}
with no $\frac12 h(0)$ boundary term: the midpoint offset
of the spectrum cancels it. Applied to
$h(x)=[e^{\beta(x-\mu)}+1]^{-1}$, for which
$h'(0)=-\beta f_0(1-f_0)$ with $|h'(0)|\le\beta/4=1/(4\tau N)$,
this gives
\begin{equation}
N=N\,\tau\log\left(1+e^{\xi/\tau}\right)
+\mathcal O\left(\frac{1}{\tau N}\right),
\label{eq:number_eq_EM}
\end{equation}
i.e.\ the continuum number equation \eqref{eq:xi_eq} is violated
only at absolute order $1/N$, and the scaled chemical potential is
shifted by $\delta\xi=\mathcal O(N^{-2})$ only. Propagating
$\delta\xi$ through the scaling functions shifts the contact by
$N^{5/2}\,\partial_\xi A\;\delta\xi=\mathcal O(N^{1/2})$. Had the
spectrum lacked the midpoint structure (e.g.\ for
$\varepsilon_n=n$), the boundary term $\frac12h(0)=\mathcal O(1)$
would have produced $\delta\xi=\mathcal O(N^{-1})$ and hence a
genuine $\mathcal O(N^{3/2})$ shift of the contact: the midpoint
structure of the harmonic spectrum is what protects the coefficient
$B(\tau)$. The same argument applies to the continuum replacements
\eqref{eq:kappa2_scaling} and \eqref{eq:kappa3_scaling} of the
cumulant sums \eqref{eq:cumulants_discrete}: their absolute
Euler--Maclaurin errors are $\mathcal O(1/N)$, i.e.\ relative
errors $\mathcal O(N^{-2})$, which perturb the fluctuation term of
\eqref{eq:FN_saddle}, itself of order $N^{3/2}$ in the
contact, only at order $N^{1/2}$.

% ===============================================================
\subsection{Wigner--Kirkwood corrections to the phase-space Fermi factor}
\label{app:pc_wk}

Equation \eqref{eq:phase_space_f} replaces the Wigner symbol of
$f_z(\hat h)$ by its classical limit. The expansion of the
symbol of a function of $\hat h=\frac{p^2}{2}+V(x)$ contains only
even powers of $\hbar$ and has the structure
\cite{BrackBhaduri2003}
\begin{equation}
\left[f_z(\hat h)\right]_W(x,p)
=f_z(H)
+\hbar^{2}\left[c_1\,f_z''(H)\,V''(x)
+c_2\,f_z'''(H)\left(V'(x)^{2}+p^{2}V''(x)\right)\right]
+\mathcal O(\hbar^{4}),
\label{eq:moyal_structure}
\end{equation}
with $H=\frac{p^2}{2}+V(x)$ and pure numbers $c_{1,2}$ whose
values are not needed here. Each derivative of $f_z$ carries a
factor $\beta=1/(\tau N)$, while in scaled variables $V''=1$ and
$V'^{\,2}=x^{2}=2Nu^{2}$: the two bracketed terms are of relative
order $1/(\tau^{2}N^{2})$ and $1/(\tau^{3}N^{2})$, respectively.
At fixed $\tau>0$ the Wigner--Kirkwood corrections to the local
contact density are therefore relative $\mathcal O(N^{-2})$
effects, contributing $\mathcal O(N^{1/2})$ to the contact. Their
$\tau$-dependent coefficients grow as $\tau\to0$ and reach
$\mathcal O(1)$ at the Airy crossover
$\tau\sim N^{-2/3}$, consistently delimiting the validity window
\eqref{eq:validity_window} of the fixed-$\tau$ expansion.
Oscillatory (Friedel-type) corrections, not captured by
\eqref{eq:moyal_structure}, are damped by thermal smearing: at
fixed $\tau$ their amplitude is exponentially small in
$T/\hbar\omega=\tau N$.

Writing the kernel through its Wigner symbol,
$K_z(x,y)=\int\frac{dp}{2\pi}\,e^{ip(x-y)}\,W_z\left(p,\frac{x+y}{2}\right)$,
the local quantities \eqref{eq:local_defs} obey the exact
relations
\begin{equation}
\rho_z=\int\frac{dp}{2\pi}\,W_z,
\qquad
S_z(x)=\frac12\,\rho_z'(x),
\qquad
\kappa_z(x)=\int\frac{dp}{2\pi}\,p^{2}\,W_z(p,x)+\frac14\,\rho_z''(x),
\label{eq:kernel_exact_identities}
\end{equation}
where the identity for $S_z$ uses the vanishing of the $p$-odd part
of $W_z$ (guaranteed by the symmetry and reality of the kernel,
$K_z(x,y)=K_z(y,x)\in\mathbb R$) and also follows directly from
$2S_z=\partial_x K_z(x,x)$. Equation \eqref{eq:S_zero} is thus a
leading-order statement whose completion is
$S_z=\frac12\rho_z'$. In scaled units $\rho_z=\mathcal O(\sqrt N)$
and $\partial_x=\mathcal O(N^{-1/2})$, so
$S_z=\mathcal O(1)$ and $S_z^{2}=\mathcal O(1)$: a relative
$N^{-2}$ correction to $\rho_z\kappa_z=\mathcal O(N^{2})$, feeding
the contact only at $\mathcal O(N^{1/2})$. Likewise
$\frac14\rho_z''=\mathcal O(N^{-1/2})$ is a relative
$\mathcal O(N^{-2})$ correction to
$\kappa_z=\mathcal O(N^{3/2})$. Together with the corrections to $W_z$ itself
(Appendix~\ref{app:pc_wk}), all centre-gradient corrections to
\eqref{eq:rho_phase_space}--\eqref{eq:kappa_phase_space} are of
relative order $N^{-2}$.

The claim \eqref{eq:CGCE_full}, that the grand-canonical contact at
fixed mean particle number contains no $N^{3/2}$ term, now follows
from the same estimates. The GCE functional \eqref{eq:FGCE_kernel}
is evaluated at a fugacity fixed by the discrete equation
$\langle N\rangle_{\GCE}=\sum_n f_n(z)=N$. Its continuum replacement induces
$\delta\xi=\mathcal O(N^{-2})$ and a contact shift of order
$N^{1/2}$. The semiclassical evaluation of
$\rho_z\kappa_z-S_z^{2}$ carries the corrections of
Appendix~\ref{app:pc_wk}, all of
relative order $N^{-2}$. Since the GCE expression contains no
contour integral, there is no Gaussian fugacity fluctuation, and no
source of a relative-$1/N$ term remains:
$\mathcal C^{\GCE}_{\langle N\rangle=N}
=A(\tau)\,N^{5/2}+\mathcal O(N^{1/2})$, as stated in
\eqref{eq:CGCE_full}. Consequently the $N^{3/2}$ term of
the canonical contact is the ensemble correction identified in
\eqref{eq:DeltaC}.

% ============================================================
\section{Numerical procedure}
% ============================================================
\label{app:numerical}
\label{sec:numerics}
Our goal here is to provide details on the numerical procedures to compute the contact from eq. \eqref{eq:contact_FN}.
We begin by defining the matrix
\begin{equation}
  R_{ab}
  = \int_{-\infty}^{\infty} dx\,
    \left[\phi_a(x)^2\,\phi_b'(x)^2
          - \phi_a(x)\,\phi_a'(x)\,\phi_b(x)\,\phi_b'(x)\right]\,,
  \label{eq:Rmatrix}
\end{equation}
truncated to $M$ single-particle levels ($a,b = 0,\ldots,M{-}1$) and
evaluated by the trapezoidal rule on a uniform grid of
$N_x = \max(4000,\,6M)$ points spanning $[-L,\,L]$ with
$L = \sqrt{2M} + 6$.  The wave functions $\phi_n(x)$ and their
derivatives are generated via the three-term recurrence
\begin{equation}
  \phi_{n+1}(x) = \sqrt{\frac{2}{n{+}1}}\;x\,\phi_n(x)
  - \sqrt{\frac{n}{n{+}1}}\;\phi_{n-1}(x)\,,
  \qquad
  \phi_n'(x) = \sqrt{\frac{n}{2}}\;\phi_{n-1}(x)
  - \sqrt{\frac{n{+}1}{2}}\;\phi_{n+1}(x)\,.
\end{equation}
With the arrays $\phi_a(x_i)$ and $\phi_a'(x_i)$ precomputed, $R_{ab}$
is assembled as $R = P - Q$ via dense matrix products of cost $\mathcal O(M^2 N_x)$.

Two numerical safeguards matter at the upper end of the
$(N,\tau)$ window. First, at large $|x|$ the recurrence seed
$\phi_0(x)=\pi^{-1/4}e^{-x^{2}/2}$ underflows in double precision
for $|x|\gtrsim\sqrt{2\times745}\simeq38.6$, which would silently
zero every $\phi_n(x)$ beyond that radius, including levels
$n\gtrsim750$ whose classically allowed region extends much
further, and bias the contact low by a factor
$\simeq\mathrm{erfc}\left(38.6/\sqrt{\tau N}\right)$ once the
thermal cloud extends past the underflow radius (a $1.5\%$ effect
at $N=50$, $\tau=10$). All high-temperature data reported here
therefore evaluate the recurrence in log-scaled form, carrying a
per-point scale exponent alongside the mantissa. Second, the
highest retained orbitals oscillate on the scale $\pi/\sqrt{2M}$,
and the quadrature grid must resolve it. The high-$\tau$ data use
$N_x=12M$ grid points.

The canonical contact is expressed as a ratio of contour integrals,
\begin{equation}
  \mathcal C_N = \frac{2}{\pi}\;\mathrm{Re} \left[
    \frac{\displaystyle \frac{1}{N_\theta}
      \sum_{k=0}^{N_\theta-1} W_k\,G_k\,e^{-iN\theta_k}}
    {\displaystyle \frac{1}{N_\theta}
      \sum_{k=0}^{N_\theta-1} W_k\,e^{-iN\theta_k}}
  \right],
  \label{eq:DFT}
\end{equation}
with $\theta_k = 2\pi k/N_\theta$, contour points $z_k = r\,e^{i\theta_k}$, and
\begin{equation}
  W_k = \exp \left[\log\Xi(z_k) - \log\Xi(z_0)\right],
  \qquad
  G_k = \sum_{a,b=0}^{M_{\mathrm{use}}-1} \bar{n}_a(z_k)\,\bar{n}_b(z_k)\,R_{ab}\,.
  \label{eq:WkGk}
\end{equation}
Here $\Xi(z) = \prod_{n=0}^{M-1}(1+z\,e^{-\beta\varepsilon_n})$,
$\bar{n}_a(z) = z\,e^{-\beta\varepsilon_a}/(1+z\,e^{-\beta\varepsilon_a})$,
and the saddle-point radius $r = e^{\beta\mu}$ is determined by
$\sum_n \bar{n}_n(r) = N$. The number of contour points
$N_\theta$ is chosen large enough that doubling it changes
$\mathcal C_N$ in \eqref{eq:DFT} by less than $10^{-5}$ in relative
units; in practice $N_\theta = 256$ is sufficient for $N\le 100$
across the $\tau\in[0,10]$ window.

Crucially, the partition function $\Xi(z)$ and the saddle-point condition
use all $M_{\mathrm{full}} = N + 14\,\tau N + 20$ levels needed for
convergence (the coefficient $14$ corresponds to retaining levels with
Boltzmann weight as low as $e^{-14}\sim 10^{-6}$ at the highest
relevant temperature; the additive offset $20$ handles the low-$\tau$
band edge), while the spatial integral $G_k$ uses only
$M_{\mathrm{use}} = \min(M_{\mathrm{full}},\,M_{\max})$ levels, where
$M_{\max}$ is the size of the precomputed $R$ matrix.
This ensures the calculation remains in the canonical
ensemble---no grand-canonical fallback is used.

\subsubsection{Spatial truncation at high $\tau$}
The spatial truncation in $G_k$ deserves careful comment, because its
behaviour at high $\tau$ is more subtle than a naive per-level estimate
suggests. A first-pass bound on the truncation error is the occupation
of the highest retained level at the saddle, $\bar n_{M_{\max}}(r)$. At
the worst corner $(N=100,\,\tau=10)$ with $M_{\max}=4000$ this gives
$\bar n_{M_{\max}}\sim 10^{-3}$, suggesting a relative deviation of order
$10^{-3}$. The actual truncation error at that corner is closer to
$15\%$. The discrepancy arises because the spatial matrix elements
$R_{ab}$ in \eqref{eq:Rmatrix} grow as $\sqrt{ab}$ for large indices, so
the contribution of pairs $(a,b)$ with $a,b\gtrsim M_{\max}$ is
 {enhanced} by a factor $\sqrt{ab}$ relative to the bare occupation,
and the number of contributing pairs is $\sim(M_{\mathrm{full}}-M_{\max})^2$.
The cumulative truncation therefore scales linearly in
$1-M_{\max}/M_{\mathrm{full}}$.

The empirically clean criterion for $G_k$ to reproduce the full-level
result to better than $\sim 0.05\%$ is
\begin{equation}
  M_{\max}\;\geq\;1.2\,M_{\mathrm{full}}(N,\tau)
  \;=\;1.2\,(N+14\tau N+20).
  \label{eq:MMAX_safety}
\end{equation}
Below this threshold the truncation bias is monotonic and well-modelled
by $\mathcal C_N^{(M_{\max})}/\mathcal C_N^{(\infty)}\simeq 1-c_\tau(1-M_{\max}/M_{\mathrm{full}})$
with $c_\tau\sim \mathcal O(1)$ and weakly $\tau$-dependent. Crossing
$M_{\max}=M_{\mathrm{full}}$ produces the steep transition to the converged
value: at $(N=20,\,\tau=10,\,M_{\mathrm{full}}=2840)$, raising $M_{\max}$
from $2000$ ($M_{\max}/M_{\mathrm{full}}=0.70$) to $2840$ removes a
truncation deficit of $\simeq0.02\%$, beyond which
$\mathcal C_N^{(M_{\max})}$ is stable at the converged value
$979.21$ (underflow-safe evaluation; stable to better than
$10^{-5}$ against further increases of $M_{\max}$ and refinements
of the quadrature grid). The prediction of the scaling law
\eqref{eq:scaling_law}, $979.27$, differs from the converged value
by $0.006\%$, consistent with the expected $\mathcal O(N^{-1})$
subleading corrections to the scaling law.

This $\tau$-dependent saturation is the reason the high-$\tau$ ratio plot
behaves differently from the low-$\tau$ one even at fixed $M_{\max}$.
At $\tau=0.05$ and $N=100$, $M_{\mathrm{full}}=190$, so any
$M_{\max}\gtrsim 250$ saturates the criterion \eqref{eq:MMAX_safety} and
the plot collapses cleanly to unity. At $\tau=10$ and the same $N$,
$M_{\mathrm{full}}=14{,}120$, so $M_{\max}=4000$--$5000$ is well below
threshold and the truncation bias dominates the apparent residual. We
therefore restrict the high-$\tau$ verification plot to the
$(N,\tau)$ region satisfying \eqref{eq:MMAX_safety} with margin
$\geq1.2$ for the chosen $M_{\max}$. There the ratio $\mathcal R_N$
defined in \eqref{eq:RN_def} collapses to unity to within $0.06\%$
for all $N\in\{10,\dots,50\}$ and $\tau\in[5,10]$, consistent with
a small next-to-subleading $N^{1/2}$ coefficient in the Boltzmann
regime. Extending the plot to higher $N$ at fixed high $\tau$
requires increasing $M_{\max}$ in proportion to $\tau N$.

\subsection{Scaling verification}
We verify the scaling ansatz \eqref{eq:scaling_law} by comparing
the numerical data against the asymptotic predictions in the low- and
high-$\tau$ regimes, and against the exact numerical evaluation of the
integral representations at intermediate~$\tau$.
Figures~\ref{fig:contact_data}, \ref{fig:full_tau}, and
\ref{fig:pade_ratio} were generated with $M_{\max} = 5000$
single-particle levels, which satisfies the safety criterion
\eqref{eq:MMAX_safety} on the windows they display. The
verification data of Figs.~\ref{fig:low_tau} and
\ref{fig:high_tau} were computed with $M_{\max}$ chosen per point
to satisfy \eqref{eq:MMAX_safety} with margin $\geq1.2$ (up to
$M_{\max}=8600$ at the largest $\tau N$) and, at high $\tau$, with
the evaluation methods described above. The corner
$(N=100,\tau=10)$ would require $M_{\max} \approx 17{,}000$ for the
same accuracy, and the high-$\tau$ window is accordingly restricted
to $N\le50$ (see the discussion below
eq.~\eqref{eq:MMAX_safety}).

The scaling functions $A(\tau)$ and $B(\tau)$ are computed from
\eqref{eq:A_tau_def} and \eqref{eq:Bsad_oneline}, respectively, with
the Fermi factor \eqref{eq:f_tau} evaluated at the self-consistent
scaled chemical potential $\xi(\tau)$ determined by \eqref{eq:xi_eq}.

The closed asymptotic forms used to anchor the verification are: for
$A(\tau)$, the Sommerfeld expansion \eqref{eq:A_tau_lowtau} at low
$\tau$ and the virial expansion \eqref{eq:A_hightau_final} at high
$\tau$; for $B(\tau)$, the Sommerfeld result
\eqref{eq:Bsad_lowtau_final_closed} at low $\tau$ and the virial
result \eqref{eq:Bsad_hightau_final_1overTau} at high $\tau$. At
intermediate $\tau\sim 1$, the integrals \eqref{eq:A_tau_def} and
\eqref{eq:Bsad_oneline} are evaluated numerically using adaptive
Gauss--Kronrod quadrature for the inner $q$-integrals and
Gauss--Legendre quadrature for the outer $u$-integral.

% ============================================================
\subsection{Pad\'e approximants}
\label{sec:pade}

Both scaling functions are smooth and monotonic in $\tau$, tend to
finite values at $\tau=0$, and grow as $\sqrt\tau$ at large $\tau$.
This structure is naturally represented by rational functions of
$\sigma:=\sqrt\tau$, and we adopt that variable throughout this
subsection. The approximants below are constructed so that
every asymptotic property established in
Appendix~\ref{app:asymptotics} is imposed: the value and the leading Sommerfeld
coefficient at $\tau\to0$, the absence of half-integer powers in the
low-temperature expansions, and the leading Boltzmann asymptote at
$\tau\to\infty$. Only the crossover region is fitted.

\subsubsection{Leading coefficient}

For $A(\tau)$ we write
\begin{equation}
  A_{\mathrm{P}}(\tau)
  = a_0
  + \frac{a_2\,\tau^{2}+\dfrac{q_4}{\pi^{3/2}}\,\tau^{5/2}}
         {1+q_1\sqrt\tau+q_2\,\tau+q_3\,\tau^{3/2}+q_4\,\tau^{2}},
  \label{eq:A_pade}
\end{equation}
with $a_0=128\sqrt2/(45\pi^{3})$ and $a_2=4\sqrt2/(9\pi)$ the
Sommerfeld coefficients of \eqref{eq:A_tau_lowtau}. In this form the constraints are
visible directly: as $\tau\to0$ the
denominator tends to unity and
$A_{\mathrm{P}}=a_0+a_2\tau^{2}+\mathcal O(\tau^{5/2})$, reproducing
\eqref{eq:A_tau_lowtau} including the absence of a linear term; as
$\tau\to\infty$ the ratio tends to
$(q_4/\pi^{3/2})\tau^{5/2}/(q_4\tau^{2})=\sqrt\tau/\pi^{3/2}$,
reproducing the leading virial asymptote of
\eqref{eq:A_hightau_final}. The four parameters $q_i$ are free and
are obtained by minimax fit (minimising the maximum relative error)
to the numerical evaluation of \eqref{eq:A_tau_def}:
\begin{equation}
  q_1 = 0.07615\,,\quad
  q_2 = 1.33429\,,\quad
  q_3 = 2.12047\,,\quad
  q_4 = 1.31842\,.
  \label{eq:A_pade_fitted}
\end{equation}
All $q_i$ are positive, so the denominator of \eqref{eq:A_pade} has
no zero on $\tau\ge0$ and the approximant is pole-free. The
resulting maximum relative error is
\begin{equation}
  \max_{\tau\in[0,10]}
  \left|\frac{A_{\mathrm{P}}(\tau) - A(\tau)}{A(\tau)}\right|
  \simeq 0.24\%,
  \label{eq:A_pade_error}
\end{equation}
with the worst case in the crossover region $\tau\sim1$.

\subsubsection{Subleading coefficient}

For $B(\tau)$ we use the same variable and write
\begin{equation}
  B_{\mathrm{P}}(\tau)
  = -\frac{\sigma}{\pi^{3/2}}\;
    \frac{a_1\,\sigma\left(1 + d_1\sigma + d_2\sigma^{2}
          + d_3\sigma^{3}\right) + d_5\,\sigma^{5}}
         {1 + d_1\sigma + d_2\sigma^{2} + d_3\sigma^{3}
          + d_4\sigma^{4} + d_5\sigma^{5}}\,,
  \qquad \sigma=\sqrt\tau,
  \label{eq:B_pade}
\end{equation}
with the single analytic coefficient
\begin{equation}
  a_1 = -\pi^{3/2}\,b_1 = \frac{16\sqrt{2}}{3\pi^{3/2}}\simeq 1.35453\,,
  \label{eq:B_pade_constraints}
\end{equation}
fixed by the Sommerfeld slope of
\eqref{eq:Bsad_lowtau_final_closed}. The structure of
\eqref{eq:B_pade} carries the low-temperature result of
Sec.~\ref{sec:Bsad_lowtau}. Because the bracket in the
numerator coincides with the first four terms of the denominator,
the numerator/denominator ratio equals $a_1\sigma+\mathcal
O(\sigma^{5})$, so
\begin{equation}
  B_{\mathrm{P}}(\tau)=b_1\,\tau+\mathcal O(\tau^{3}),
  \label{eq:B_pade_lowtau_structure}
\end{equation}
i.e.\ the approximant carries no $\tau^{3/2}$ term and
no $\tau^{2}$ term, the structure established by
 \eqref{eq:edge_cancellation}, with the first
correction appearing at $\mathcal O(\tau^{3})$.  At large
$\sigma$ numerator and denominator are both dominated by
$d_5\sigma^{5}$, so
$B_{\mathrm{P}}\to-\sigma/\pi^{3/2}=-\sqrt\tau/\pi^{3/2}$, the
leading Boltzmann asymptote of
\eqref{eq:Bsad_hightau_final_1overTau}. The five free parameters are
determined by minimax fit to the numerical evaluation of
\eqref{eq:Bsad_oneline} on $\tau\in[0.01,10]$:
\begin{equation}
\begin{aligned}
  &d_1 = 0.40774\,,\quad d_2 = 0.43708\,,\quad d_3 = 2.44382\,,\\
  &d_4 = 2.95067\,,\quad d_5 = 7.28346\,.
\end{aligned}
  \label{eq:B_pade_fitted}
\end{equation}
As for $A_{\mathrm{P}}$, all $d_i$ are positive and the approximant
is pole-free on $\tau\ge0$. The maximum relative error is
\begin{equation}
  \max_{\tau\in[0.01,\,10]}
  \left|\frac{B_{\mathrm{P}}(\tau) - B(\tau)}{B(\tau)}\right|
  \simeq 0.26\%.
  \label{eq:B_pade_error}
\end{equation}
Both approximants remain accurate outside the fitting window, since
they are asymptotically correct by construction: at $\tau=100$ one
finds $A_{\mathrm{P}}\simeq1.7928$ and
$B_{\mathrm{P}}\simeq-1.8000$, against the leading asymptotes
$\pm\sqrt{100}/\pi^{3/2}\simeq\pm1.7959$. Equations
\eqref{eq:A_pade} and \eqref{eq:B_pade} are therefore uniformly
valid closed-form substitutes for the universal integrals
\eqref{eq:A_tau_def} and \eqref{eq:Bsad_oneline} across the full
physical temperature range.

In analogy with the ratio $\mathcal R_N$ defined in \eqref{eq:RN_def},
we introduce its Pad\'e counterpart
\begin{equation}
  \mathcal R_N^{(\mathrm{P})}(\tau)
  := \frac{\mathcal C_N(\tau)}
          {A_{\mathrm{P}}(\tau)\,N^{5/2}+B_{\mathrm{P}}(\tau)\,N^{3/2}},
  \label{eq:RNP_def}
\end{equation}
in which the universal scaling functions $A(\tau)$ and $B(\tau)$ are
replaced by their Pad\'e approximants $A_{\mathrm{P}}(\tau)$ and
$B_{\mathrm{P}}(\tau)$ from \eqref{eq:A_pade} and \eqref{eq:B_pade}.

\bibliographystyle{apsrev4-2}
\bibliography{refs}
%======================================================================

\end{document}